\documentclass[10pt,journal,compsoc]{IEEEtran}
\usepackage{graphicx}
\usepackage{amsmath}
\usepackage{multirow}
\usepackage{mathtools}
\usepackage{placeins}
\usepackage{float}
\usepackage{booktabs}
\usepackage{ragged2e}
\usepackage{xcolor}

\usepackage{amsfonts}
\usepackage{xcolor}

\usepackage{graphicx}
\usepackage{amsmath}
\usepackage{multirow}
\usepackage{mathtools}
\usepackage{placeins}
\usepackage{float}
\usepackage{booktabs}
\usepackage{ragged2e}

\ifCLASSOPTIONcompsoc
  \usepackage[nocompress]{cite}
\else
  \usepackage{cite}
\fi

\begin{document}
\title{Duration modeling with semi-Markov Conditional Random Fields for keyphrase extraction}

\author{Xiaolei~Lu,Tommy~W.S.Chow,~\IEEEmembership{Fellow,~IEEE}%

\IEEEcompsocitemizethanks{\IEEEcompsocthanksitem Xiaolei Lu is with
the Dept of Electronic Engineering at the City University of Hong
Kong, Hong Kong (Email:xiaoleilu2-c@my.cityu.edu.hk)
\IEEEcompsocthanksitem Tommy W S Chow is with the Dept of Electronic
Engineering at the City University of Hong Kong, Hong Kong (E-mail:
eetchow@cityu.edu.hk).}}

\IEEEtitleabstractindextext{%

\begin{abstract}
Existing methods for keyphrase extraction need preprocessing to generate candidate phrase or post-processing to transform keyword into keyphrase. In this paper, we propose a novel approach called duration modeling with semi-Markov Conditional Random Fields (DM-SMCRFs) for keyphrase extraction. First of all, based on the property of semi-Markov chain, DM-SMCRFs can encode segment-level features and sequentially classify the phrase in the sentence as keyphrase or non-keyphrase. Second, by assuming the independence between state transition and state duration, DM-SMCRFs model the distribution of duration (length) of keyphrases to further explore state duration information, which can help identify the size of keyphrase. Based on the convexity of parametric duration feature derived from duration distribution, a constrained Viterbi algorithm is derived to improve the performance of decoding in DM-SMCRFs. We thoroughly evaluate the performance of DM-SMCRFs on the datasets from various domains. The experimental results demonstrate the effectiveness of proposed model.

\end{abstract}

}


\maketitle

\IEEEdisplaynontitleabstractindextext

%
\IEEEpeerreviewmaketitle

\IEEEraisesectionheading{\section{Introduction}\label{sec:introduction}}

The past few decades of human history have witnessed the surge of huge amounts of text data appeared in electronic forms, such as digital libraries and social media. The drifting of publication format from hard print to digital is changing our reading habit. The reading culture of human being has gradually moved from paper to electronic platform. Identifying and retrieving interested and required information from massive text corpora rapidly and accurately is critical for the highly competitive information age \cite{re1}. Keyphrases usually best describe the main idea of the document. The combination of a few keyphrases can be treated as a concentration of the whole document, which greatly benefits information retrieval and other natural language processing (NLP) tasks, for example, literature retrieval \cite{re2}, document summary \cite{re3} and document classification \cite{re4}.

Most research articles have keyphrases assigned by authors, but a large fraction of documents only include titles and contexts. Keyphrase extraction improves the efficiency of indexing and provides a brief summarization of the document to readers. The challenge of keyphrase extraction lies in the ways of how the importance of words in the document is determined and measured. For instance, the relativeness between words and major theme of a given document is a widely used metric. Generally, the main idea of keyphrase extraction comes from the global structure of a document.  Apparently, accurately extracting global features of words (e.g. the dependency relationship with surrounding words) is essential in identifying keyphrases.

To date, the unsupervised methods used for keyphrase extraction mainly focus on score ranking. Typically, statistical feature TFIDF (term frequency inverse document frequency) \cite{re5} is widely used to extract the words that frequently occur in a document but not frequently in the whole collected documents. Another effective way is graph-based ranking algorithms \cite{re6}. For example, given a document, we can build a graph in which nodes represent different words and edges describe syntactic and/or semantic relationship between two words. Then graph rank algorithm is applied to compute score of each node, and nodes with higher scores are more important. However, most unsupervised methods require post-processing to form keyphrases as the ranking is based on word-level. Also, the evaluation results rely on the selection of top K words from ranking \cite{re7}. 

Supervised methods consider keyphrase extraction as a binary classification task. The most common way is to generate candidate phrases which are subsequently classified into keyphrase or non-keyphrase. The major concern of supervised methods is feature design.  The features of a candidate phrase can be extracted from training documents and external resources (e.g. Wikipedia articles). For example, part-of-speech (POS) tags assigned to the keyphrase and the number of Wikipedia articles where the candidate appears. Since hand-craft features are domain-specific, feature representation learning \cite{re8} has been exploited to improve the generalization. 

The selection of candidate phrases (or words) is fundamental to the above methods as candidate phrases (or words) are the input of ranking and classification models \cite{re9}. The commonly selection criteria is to generate n-grams based on the prior knowledge about keyphrases. For example, selecting the phrases containing at most three words or selecting noun phrases \cite{re10}. However, the number of n-grams can be very large for a long document. Different filtering conditions, such as the lower bound of frequency of phrase \cite{re11} and the low rectified frequencies \cite{re12}, have been designed to generate high quality candidate phrases.

In recent years keyphrase extraction is reformulated as sequence labeling to avoid the process of generating candidate phrases. In sequence labeling, a document or a sentence can be viewed as a sequence and each element (word) in the sequence is tagged with keyword or non-keyword. A keyphrase is then formed by combining consecutive keywords. Conditional random fields (CRFs) have been widely adopted for keyphrase extraction \cite{re13} \cite{re14}. When encoding the phrase-level feature "isNounPhrase" that is important in identifying keyphrases, however, CRFs fail to capture multiple phrase-level information. For example,  "a", "lot", "of" and "attention" are all tagged with NounPhrase for CRFs \cite{re13}. But based on the analysis of parse tree in Figure 1, "of attention" should not be a noun phrase in terms of two-words phrase-level. 

\begin{figure}
\centering
\scalebox{0.6}{
\includegraphics[width=5in,height = 3in ]{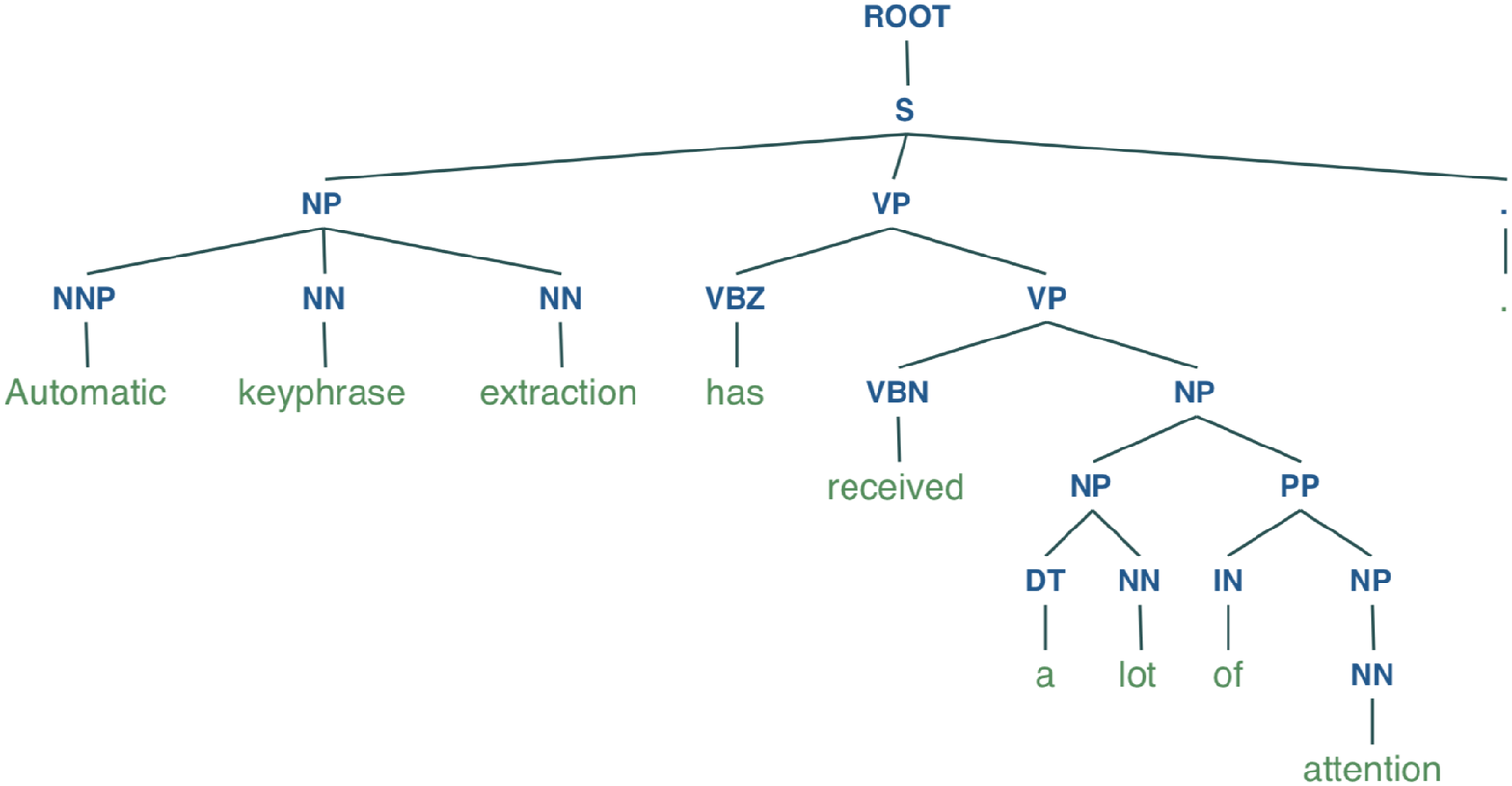}}\label{fig:secondfigure}
\caption {Parse tree analysis} 
\end{figure}

There are certain research work focusing on Semi-Markov conditional random fields (Semi-CRFs) to generalize CRFs with semi-Markov chain. Figure 2 shows a simple semi-Markov chain, in which each state has a variable duration $d$ (i.e. sojourn time) rather than a unit length of time. Extending semi-Markov chain to CRFs, the duration can be measured at the segments with various lengths which refer to the number of observations. For example, as shown in Figure 4, duration of the state "$y_{1}$" is 2. Semi-CRFs were proposed to allow modeling different phrase-level features within the boundary of phrases. As shown in Table 1, let the maximum length of a phrase is 4, the feature "isNounPhrase" for "attention",  "of attention" and "lot of attention" can be extracted, where NP refers noun phrase and N-NP is non-noun phrase. Semi-CRFs have been successfully applied in noun phrase (NP) chunking \cite{re15}, named entity recognition (NER) \cite{re16} and opinion extraction \cite{re17}. In these tasks, Semi-CRFs do not consider duration information which is effective in distinguishing segments. Take speech recognition as an example, speech duration helps to distinguish words in English [18], such as \textit{sit} and \textit{seat}. Furthermore, some research work only focuses on efficient training algorithm for Semi-CRFs as the inference complexity is proportional to the maximum length $L$ [19] [20], whilst efficient decoding to find the best segmentation with label assignment receives much less attention.

\begin{figure}
\centering
\scalebox{0.9}{
\includegraphics[width=3in,height = 1in ]{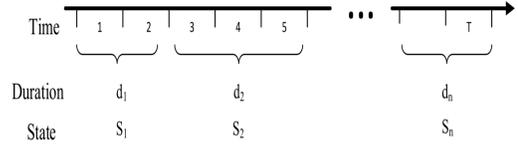}}\label{fig:secondfigure}
\caption {A semi-Markov chain.} 
\end{figure}

When applied Semi-CRFs to keyphrase extraction, compared with CRFs, Semi-CRFs allow constructing features of phrases in different length ${1,...,L}$ and do not need post-processing to generate keyphrase. Table 2 shows the difference of label tagging between Semi-CRFs and existing supervised methods for keyprase extraction. Semi-CRFs sequentially classify phrases as KP (keyphrase) or NKP (non-keyphrase) rather than sequentially classify words. This mechanism is more appropriate for extracting keyphrases.

In this paper, to address the above two problems in Semi-CRFs, through exploiting the characteristics of Semi-CRFs we derive duration modeling with semi-Markov conditional random fields (DM-SMCRFs) to further explore duration information of keyphrases. The results are significant in improving the performance of keyphrase extraction    because of more accurate keyphrase segmentation and effective decoding. Our contributions can be summarized as follows:

First, by assuming the independence between state transition and state duration, DM-SMCRFs model the distribution of duration of keyphrase to further explore state duration information. Since the keyphrase is more likely to be in a form of specific number of words (e.g. two words), explicitly modeling the duration of keyphrase can help distinguish the size of keyphrase. As shown in Table 2, the phrase "multiple autonomous sources" and "autonomous sources" have the same or similar features, but the model is expected to tag "autonomous sources" as KP. Gaussian and Gamma distributions are employed to model the duration of keyphrase as they are the good approximations of the empirical distribution of kephrase's duration.  And we further investigate which is the best fit in terms of performance of kepyrase extraction.

Second, the constrained Viterbi algorithm is derived to improve the effectiveness of decoding in DM-SMCRFs. Since most of manually assigned keyphrases are noun phrases, the hard constraint that non-noun phrase can't be tagged as keyphrase is incorporated into the decoding process. Then based on the convexity of parametric duration feature derived from duration distribution, subpaths that have no chance to result in the best predecessor of state KP can be pruned out. The above two constraints effectively reduce the average number of possible transitions between segments. Additionally, the proposed hard constraint corrects the wrong assignment of labeling non-noun phrase as KP, which slightly improves the performance of keyphrase extraction. 

Third, in order to demonstrate the effectiveness of the proposed model, we collect datasets from various domains, such as Psychology, Economics and History. The experimental results show that our proposed approach outperforms the traditional methods.

\begin{table}
\centering
\caption{Phrase-level feature of semi-CRFs}
\begin{tabular}{ll} 
\hline
Subsequence        & isNounPhrase  \\ 
\hline
attention          & NP           \\
of attention       & N-NP         \\
lot of attention   & NP           \\
a lot of attention & NP           \\
\hline
\end{tabular}
\end{table}

\begin{table*}[]

\centering \caption{Keyphrase extraction using semi-CRFs and
traditional methods} \label{my-label}
\scalebox{0.95}{
\begin{tabular}{|c|c|c|c|c|c|c|c|c|c|c|c|c|}
\hline Methods & \textbf{Big}                        & \textbf{data}
& concern      & large-volume      & complex      & growing       &
data      & sets      & with     & multiple                &
\textbf{autonomous}                & \textbf{sources}
\\ \hline semi-CRFs
& \multicolumn{2}{c|}{KP} & NKP          & NKP               & NKP
& \multicolumn{3}{c|}{NKP}              & NKP      & NKP &
\multicolumn{2}{c|}{KP}
\\ \hline
\begin{tabular}[c]{@{}c@{}}CRFs\end{tabular} & KP                                  & KP                                   & NKP          & NKP               & NKP          & NKP           & NKP       & NKP       & NKP      & NKP                     & KP                                 & KP                             \\ \hline
\begin{tabular}[c]{@{}c@{}}Traditional\\ Supervised method\end{tabular} & \multicolumn{2}{c|}{\begin{tabular}[c]{@{}c@{}}Big data\\ KP\end{tabular}} & \multicolumn{7}{c|}{\begin{tabular}[c]{@{}c@{}}large complex growing data sets\\ NKP\end{tabular}} & \multicolumn{3}{c|}{\begin{tabular}[c]{@{}c@{}}multiple autonomous sources\\ KP\end{tabular}} \\ \hline
\end{tabular}}
\end{table*}

\section{Related work}
The unsupervised methods for keyphrase extraction usually contain two steps: word unit ranking and keyphrase formation. Mihalcea and Tarau \cite{re21} proposed the TextRank model which is inspired by PageRank algorithm. In this graph model, each node represent a word in the document and edge describes the co-occurrence relation of two nodes within a fixed window size. After the ranking process, a candidate phrase is chosen as a keyphrase if it includes one or more top-ranked words. In graph ranking algorithm, the score assigned to each node has four different centrality measures. Since TextRank is based on eigenvector measure, Boudin \cite{re22} provided another three different measures: degree centrality, closeness centrality and betweenness centrality. Some ranking models that are extensions of TextRank have been proposed to improve the performance of
keyphrase extraction. For example, ExpandRank \cite{re23}, using k nearest neighboring documents to facilitate co-occurrence statistics. SingleRank \cite{re23}, which selects the K
highest-scored candidate phrases as keyphrases by summing the scores of constituent words after graph ranking.

Another unsupervised method for keyphrase extraction is based on clustering approach. Assuming keyphrases usually represent various semantic topics of the document, Liu et al. \cite{re24} used three clustering methods (i.e. affinity propagation, hierarchical clustering and spectral clustering) to group candidate terms into clusters, where within-cluster terms share similar semantics. Then keyphrases are extracted from these clusters by identifying the exemplar terms. Liu et al. \cite{re25} further proposed Topical PageRank to measure importance of words by incorporating topic information. Grineva et al. \cite{re26} also proposed a topic-based clustering method to partition the term graph into groups with different topics and select the groups containing key terms. However, it should be noted that in most cases manually assigned keyphrases can not cover the main topics of the document \cite{re27}.

Supervised methods have been applied in GenEx \cite{re28} and Kea \cite{re29} automatic keyphrase extraction systems. Based on the parameters learned with decision trees in the training process, GenEx uses the genetic algorithm to adjust these parameters in order to optimize the performance on training documents. Kea employs the Naive Bayes model to compute the overall probability of a candidate phrase being a keyphrase. Then post-processing is operated to eliminate the candidate keyphrases which are subphrases of another
candidate keyphrases. Other supervised learning methods like bagging \cite{re30} and boosting \cite{re31} have been used to train a binary classifier on phrases annotated with keyphrase or non-keyphrase. In order to determine the importance of identified keyphrases, Jiang et al. \cite{re32} applied Linear Ranking SVM to rank candidate keyphrases. Besides, external resources like terminological databases \cite{re33} have been exploited to enrich the features for keyphrase extraction.

Keyphrase extraction can be formed as a sequence labeling task. Considering an observation sequence corresponding to the words in a document, a sequence of labels from the set of \{keyphrase, non-keyphrase\} is assigned to each word. Conditional random fields (CRFs) \cite{re34} have been widely used in sequence labeling as it relaxes strict independence assumption in Hidden Markov Models by directly modeling the global conditional distribution over observation sequence. ZHANG et al. \cite{re35} first applied CRFs model to extract keywords. CRFs encode local and global features of a text sequence, which can then improve the performance of keyword extraction. Gollapalli et al. \cite{re14} further enhanced extraction performance by incorporating expert knowledge in CRFs model, where adjacent words tagged with keyword forms a keyphrase. 

Semi-CRFs \cite{re16} were proposed to allow constructing segment-level features for named entity recognition (NER). Encoding segment-level features such as “entity length” and “similarity to other known entities” can help improve the performance of NER. Yang and Cardie \cite{re17} applied semi-CRFs to extract opinion expressions and proposed to incorporate parse tree information to generate meaningful segments. Hierarchical semi-Markov conditional random field (HSCRF) \cite{re36} was proposed to model complex hierarchical Markov processes, which can be used to jointly infer the tags of noun phrases and part-of-speech tags. Since semi-CRFs suffer from high computational cost compared with CRFs, Sarawagi \cite{re19} designed an efficient training algorithm to learn the features common across overlapping segments. Okanohara et al. \cite{re20} used feature forests to pack feature-equivalent states and filter process to choose candidate states. Besides, Muis et al. \cite{re15} proposed Weak semi-CRFs that determine the length of next segment and state separately, which can reduce the time complexity but fail to capture the dependency between the state and its duration.

\section{Review of CRFs and Semi-CRFs}
Conditional random fields (CRFs) were first developed by Lafferty et
al. \cite{re34} for labeling sequence data. Linear-chain CRFs have
been widely used in NLP tasks, for example, part-of-speech tagging
\cite{re37} and named entity recognition \cite{re38}. Sarawagi et
al. \cite{re16} extended CRFs to semi-Markov case, which can
incorporate segment-level features and sequentially assign labels to segments of
input sequence. In this section, we briefly review CRFs and
Semi-CRFs.
\subsection{Conditional random fields}
CRFs define a conditional probability distribution
$P\left(Y|X\right)$ over the label sequence $Y$ and observation
sequence $X$. It relaxes the independence assumption in Hidden
Markov Models (HMMs). Given a linear chain CRFs with the observation
sequence $X=\{x_{1},x_{2},...,x_{n}\}$ and corresponding label
sequence $Y=\{y_{1},y_{2},...,y_{n}\}$, which is visualized in
Figure 3, the conditional probability is defined as
\begin{equation}
  P(Y|X;\theta)=\frac{1}{Z(X)}\exp(\sum_{k=1}^{K}\theta_{k}F_{k}(X,Y)),
\end{equation}
where $F_{k}(X,Y)=\sum_{i=1}^{n}f_{k}(y_{i},y_{i-1},X,i)$ are
feature functions. $Z(X)=\sum\nolimits_{y_{i}}\exp(\sum_{k=1}^{K}\theta_{k}f_{k}(y_{i},y_{i-1},X,i))$
is the partition function and $\theta$ is a weight vector.

Given the training data includes observation sequences
$\{X^{1},X^{2},...,X^{m}\}$ and label sequences
$\{Y^{1},Y^{2},...,Y^{m}\}$, the negative log-likelihood of CRFs can
be defined as
\begin{equation}
  L(\theta)=-\sum_{i=1}^{m}logP(Y^{i}|X^{i};\theta)+\frac{\|\theta\|^{2}}{2\sigma^{2}}.
\end{equation}

We can estimate parameter $\theta$ by minimizing $L(\theta)$, in
which optimization algorithms like conjugate gradient
and L-BFGS \cite{re39} can be used.

\begin{figure}
\centering
\scalebox{0.9}{
\includegraphics[width=4in,height = 1in ]{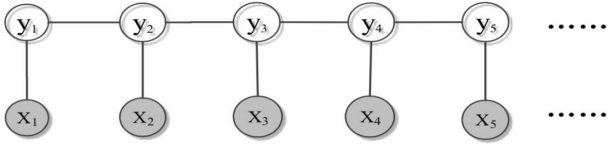}}
\caption {The graph structure of CRFs.} 
\end{figure}

\subsection{Semi-Markov conditional random fields}
Considering an observation sequence $X=\left \{x_{1},x_{2},...,x_{n}\right \}$, let
$S=\left \{s_{1},s_{2},...,s_{p}\right \}$ indicates a possible segment sequence
for $X$, where $s_{j}=(t_{j},u_{j},y_{j})$, $t_{j}$ is the start
position of segment $s_{j}$ and $u_{j}$ is the end position. All the
elements in this segment share the same label $y_{j}$. Figure 4
shows the graph structure of Semi-CRFs (gray box denotes transition factor that corresponds to state transition feature), in which state transition only depends on the current segment and the label of previous segment. Let $ \textbf{g} =\left ( g_{1} ,..,g_{K}\right ) $ is the vector of segment feature functions. Each segment $(j,X,S)$ corresponds to a feature measurement $g_{k}(j,X,S)$,  which is defined as

\begin{figure}
\centering
\scalebox{0.9}{
\includegraphics[width=4in,height = 1in ]{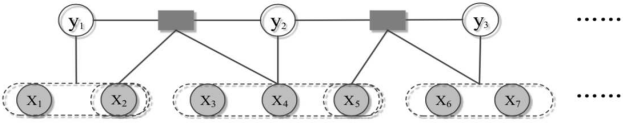}}
\caption {The graph structure of Semi-CRFs.} 
\end{figure}

\begin{equation}
g_{k}(j,X,S)=g_{k}(y_{j-1},y_{j},X,t_{j},u_{j})
\end{equation}

The conditional probability of Semi-CRFs is represented by
\begin{equation}
  P(S|X;\theta)=\frac{1}{Z(X)}\exp \left \{ \theta G(X,S)\right \},
\end{equation}
where $G(X,S) = \sum_{j=1}^{|S|}\textbf{g}(j,X,s) $, $Z(X)=\sum\nolimits_{S'}\exp \left \{  \theta G_{k}(X,S')\right \}$. $ \theta =\left ( \theta_{1} ,...,\theta_{K}\right ) $ is a weight vector.

Usually, segment length in Semi-CRFs is fixed with $L$, Equation (4) can be
rewritten as
\begin{equation}
\scalebox{0.9} {$
 P(S|X)=\frac{1}{Z(X)}\exp \left \{ \sum_{k=1}^{K}\sum_{d=1}^{L}\sum_{i=1}^{n}\theta_{k}g_{k}(y',y,X,i-d+1,i) \right \}, $}
\end{equation}
where $(i-d+1)$ denotes the start position of current segment.
$y$ represents the label of current segment while $y'$ is the
label of previous segment. $Z(X)$ is computed as
\begin{equation}
Z(X)=\sum\nolimits_{y}\exp \left \{  \sum_{k=1}^{K}\sum_{d=1}^{L}\sum_{i=1}^{n}\theta_{k}g_{k}(y',y,X,i-d+1,i)\right \}.
\end{equation}

Considering a training set $\{X^{i},S^{i}\}_{i=1}^m$, the parameter
$\theta$ can be estimated by minimizing the negative log-likelihood
defined as follows:
\begin{equation}
  L(\theta)=-\sum_{i=1}^{m}logP(S^{i}|X^{i};\theta)+\frac{\|\theta\|^{2}}{2\sigma^{2}}.
\end{equation}

\section{Proposed method}

In this section, we present the algorithm of DM-SMCRFs. It exploits the characteristics of Semi-CRFs to model segment-level features. DM-SMCRFs further explore the duration information by modeling duration distribution of keyphrase to improve the performance and efficiency of keyphrase extraction. 

\subsection{Model description}

Given a sequence $X=\left \{ x_{1},x_{2},...,x_{n}\right \}$, let $S=\left \{ s_{1},s_{2},...,s_{q} \right \}$ denotes a sequence of consecutive segments, where
$s_{j}$ is a tuple consisting of $(y_{j},t_{j},u_{j})$. $y_{j} $ is the label of the segment $s_{j}$, $t_{j}$ and $u_{j}$ represent the start and end position of $s_{j}$ respectively. The range of segment length is $[1,L]$. Figure 5 shows the graph structure of DM-SMCRFs. 

First, similar to Semi-CRFs, DM-SMCRFs assume that the state transition is independent of duration of previous state. Figure 5 shows that transition factor only depends on the last element in the previous segment. This assumption is more intuitive in keyphrase extraction as the length of phrase do not account for the change of state transition probability. For example, when tagging a sentence as shown in Table 2, the state KP for the first two words "Big data" or the second word "data" has no effect on tagging the next word "concerns" as NKP. 

Second, DM-SMCRFs additionally assume that the duration of current state is independent of previous state. As shown in Figure 5, the transition factor does not depend on the length of current segment. Figure 6 shows state duration modeling in DM-SMCRFs, where ${y}'$ and $y$ are the state of previous and current segment respectively and $p_{y}(d)$ represents the probability of staying duration $d$ in state $y$. In most cases, keyphrases are more likely to be in a form of specific number of words (e.g. two words). This assumption allows explicitly modeling the duration a phrase is expected to be in a particular state. For instance, the phrase "autonomous source" is more likely to be tagged with KP while "multiple autonomous source" is expected to be in the state NKP.

\begin{figure}
\centering
\scalebox{0.9}{
\includegraphics[width=4in,height = 1in ]{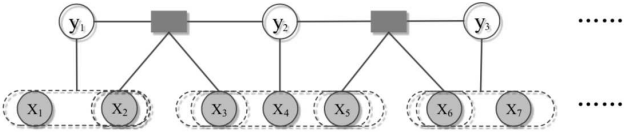}}
\caption {The graph structure of DM-SMCRFs.} 
\end{figure}

\begin{figure}
\centering
\scalebox{0.9}{
\includegraphics[width=3in,height = 1.8in ]{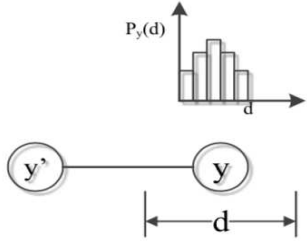}}
\caption {State duration modeling.} 
\end{figure}

Traditionally, there are two types of feature functions defined in Semi-CRFs. They are observation feature function $b_{k}(y,X,i-d+1,i)$ and state transition feature function $s_{k}({y}',y,X,i-d+1,i)$. When incorporating the above assumptions in DM-SMCRFs, the state transition feature function $s_{k}({y}',y,X,i-d+1,i)$ is decomposed into transition and duration feature function, which are defined as follows:

\begin{equation}
\left\{
\begin{array}{lcl}
Observation:b_{k}(y,X,i-d+1,i),\\
Transition:t_{{k}'}(y',y,X,i-d+1),\\
Duration:D^{y}d_{{k}''}(y,X,i-d+1,i),
\end{array} \right.
\end{equation}
where ${y}'$ and $y$ are the label of previous and current segment respectively. $D^{y}$ represents state duration feature, which refers to the measurements of state duration information and is described in Section 4.2.

These feature functions are expressed as

\begin{equation}
\left\{
\begin{array}{lcl}
b_{k}(y,X,i-d+1,i)=\mathbb{I}(y=I)\mathbb{I}(X_{(i-d+1:i)}=O ),\\

t_{{k}'}({y}',y,X,i-d+1) = \mathbb{I}(y=I)\mathbb{I}({y}'=Q) ,\\

d_{{k}''}(y,X,i-d+1,i)=\mathbb{I}(y=I)\mathbb{I}(d=l),

\end{array}  \right.
\end{equation}
where the indicator function $\mathbb{I} (F) = 1$ if $F$ if true and zero otherwise. $I$ and $Q$ denote current and previous label configuration respectively. $O$ represent the extracted feature from the segment $X_{(i-d+1:i)}$ while $l$ is the length of the segment.

The conditional probability of DM-SMCRFs is defined as 

\begin{equation}
\scalebox{0.8}{$\displaystyle {P(S|X;\theta ) =  \frac{1}{Z(X)}\exp
\left\{
\begin{array}{lcl}
\sum_{k=1}^{K_{0}}\sum_{d=1}^{L}\sum_{i=1}^{n}\theta_{k}b_{k}(y,X,i-d+1,i)+\\
\sum_{{k}'=1}^{K_{1}}\sum_{i=1}^{n}\theta _{{k}'}t_{{k}'}(y',y,X,i-d+1)+\\
\sum_{{k}''=1}^{K_{2}}\sum_{d=1}^{L}\sum_{i=1}^{n}\theta_{{k}''}D^{y}d_{{k}''}(y,X,i-d+1,i).
\end{array} \right.} $}
\end{equation}
where $Z(X)$ is the summarization of all possible segmentations for the observational sequence $X$.

\subsection{Duration modeling}

In keyphrase extraction, duration is measured at the phrases with different length (i.e. the number of words). There are two states for keyphrase extraction: KP (keyphrase) and NKP (non-keyphrase). Since non-keyphrase can have arbitrary lengths, we focus on modeling the duration of the state KP which is usually in the limited range. 

Every discipline has its own specialist terms, which can make a difference in manually assigned keyphrases. In this paper, we investigate about 2,500 research documents from different domains: Engineering, History, Economics and Psychology. Figure 7 shows the histogram of the length of keyphrase which can be denoted as the duration of state KP. We observed that the duration of state KP has a limited range of variation and each frequency distribution has one clear peak, which  is similar to the duration feature of sound segments \cite{re40}.Since Gaussian and Gamma distributions have been widely accepted to model the duration of sounds segments in speech recognition, we apply these two distributions to model the duration of state KP and further investigate which is better in terms of the performance of keyphrase extraction. By modeling the state duration with the best parametric density which is estimated with maximum likelihood (ML) \cite{re41}, we take the discrete counterpart of this density as the best probability mass function. Figure 8 displays the empirical distribution of the duration of state KP with the corresponding Gaussian and Gamma fit. It can be seen that Gamma distribution fits the Engineering, History and Economics dataset quiet well while the Gaussian fit is better than the Gamma fit in Psychology dataset.

\begin{figure}
\centering
\includegraphics[width=4in,height = 3in ]{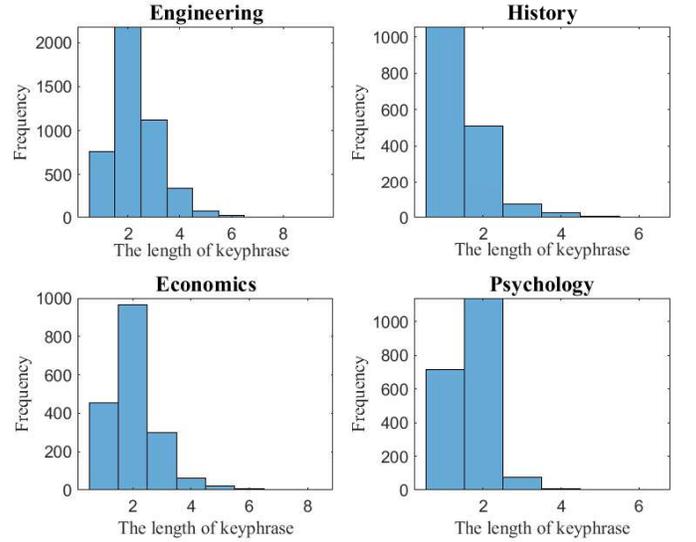}
\caption {Histogram of the length of keyphrase from various domains.} 
\end{figure}

\begin{figure}
\centering
\includegraphics[width=3.5in,height = 3in ]{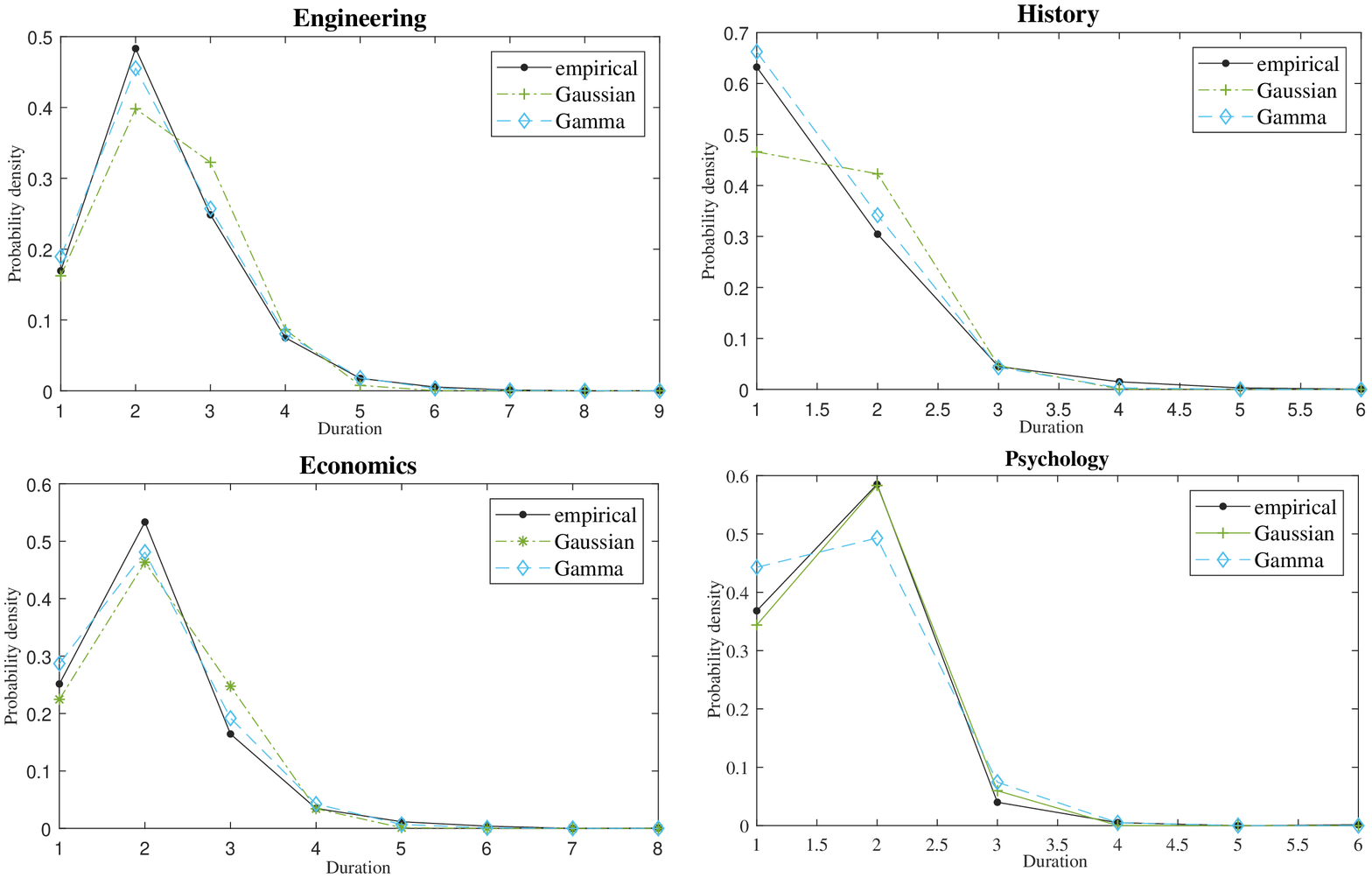}
\caption {Empirical distribution of the length of keyphrase from various domains.} 
\end{figure}

The discrete Gaussian and Gamma distribution \cite{re42} can be defined as follows:
\begin{equation}
Gaussian:p(d)=K_{1}\exp\left \{ -\frac{\left ( d-\mu  \right )^{2}}{2\sigma ^{2}}    \right \},d=0,1,2...
\end{equation}
where $K_{1}$ is a constant, $\mu$ and $\sigma$ are the mean and variance of the discrete variable.

\begin{equation}
Gamma:p(d)=K_{2}\exp\left \{ -\alpha d \right \}d^{p-1},d=0,1,2...
\end{equation}
where $K_{2}$ is a normalizing term, $\alpha$ is the shape parameter and $p$ is the rate parameter.

Therefore we define the duration feature for state KP as

\begin{equation}
\left\{
\begin{array}{lcl}

-\frac{(d-\mu )^{2}}{2\sigma ^{2}} & & {Gaussian-like},\\
-\alpha d+\beta \ln d              & & {Gamma-like},\\
\end{array} \right.
\end{equation}
where $\beta = p-1$.

When applying DM-SMCRFs to keyphrase extraction, the duration feature $D^ y$ is expressed as

\begin{equation} Gaussian \ Distribution : D^ y = \left\{
\begin{array}{lcl}
-\frac{(d-\mu )^{2}}{2\sigma ^{2}} & &   y = KP,\\

1 && y \neq KP.

\end{array} \right.
\end{equation}

\subsection{Parameter estimation}
Consider the training sequences
${(X^{1},S^{1}),(X^{2},S^{2}),...,(X^{N},S^{N})}$, $N$ is the number
of sequences, following the Semi-CRFs proposed by Sarawagi and Cohen
\cite{re16}, the negative log-likelihood over the sequences can be
written as follows:
\begin{equation}
 \begin{split}
L(\theta)  & = \sum_{q=1}^{N}P(S^{q}|X^{q};\theta )+\frac{\left\|\theta  \right \|^{2}}{2\sigma ^{2}}\\
            & = \sum_{q=1}^{N}\left \{ logZ(X^{q})-\sum_{k=1}^{K_{0}}\theta _{k}B_{k}(X^{q},S^{q})\right.\\
            &\left. - \sum_{{k}'=1}^{K_{1}}\theta _{{k}'}T_{{k}'}(X^{q},S^{q})-\sum_{{k}''=1}^{K_{2}}\theta _{{k}''}G_{{k}''}(X^{q},S^{q}) \right\} \\
            &+ \frac{\left\|\theta  \right \|^{2}}{2\sigma ^{2}},
 \end{split}
\end{equation}
where $B_{k},T_{{k}'},G_{{k}''}$ are defined as

\begin{equation}
\left\{
\begin{array}{lcl}

B_{k}=\sum_{d=1}^{L}\sum_{i=1}^{n^{q}}b_{k}(y^{q},X^{q},i-d+1,i),\\
T_{{k}'}=\sum_{i=1}^{n^{q}}t_{{k}'}(y'^{q},y^{q},X^{q},i-d+1),\\
G_{{k}''}=\sum_{d=1}^{L}\sum_{i=1}^{n^{q}}D^{y^q}d_{{k}''}(y^{q},X^{q},d).

\end{array} \right.
\end{equation}

For convenience, Equation (15) can be rewritten as
\begin{equation}
L(\theta )=\sum_{q=1}^{N}\left \{ logZ(X^{q})-\sum_{k=1}^{K}\theta
_{k}F_{k}(X^{q},S^{q}) \right \}+\frac{\left \| \theta  \right
\|^{2}}{2\sigma ^{2}},
\end{equation}
where $F_{k}(X,S)$ is the combination of observation function,
transition function and duration function.

In order to find the optimal parameter value $\theta^{*}$, firstly
we compute the derivative $\frac{\partial L(\theta )}{\partial
\theta _{k}}$ as
\begin{equation}
\frac{\partial L(\theta )}{\partial \theta _{k}}=\sum_{q=1}^{N}\left
\{ E_{P(S^{q}|X^{q};\theta )}[F_{k}(X^{q},S^{q})]-F_{k}(X^{q},S^{q})
\right \}+\frac{\theta _{k}}{\sigma ^{2}},
\end{equation}

\begin{equation}
 \begin{split}
\frac{\partial }{\partial \theta_{k}}logZ(X)
&=\frac{1}{Z(X)}\sum\nolimits_{S^{q}}\frac{\partial
Z(X^{q})}{\partial \theta _{k}}\\
&=\frac{1}{Z(X)}\sum\nolimits_{S^{q}}F_{k}(X^{q},S^{q})\exp{\sum_{k=1}^{K}\theta
_{k}F_{k}(X^{q},S^{q})}\\
&=\sum\nolimits_{S^{q}}F_{k}(X^{q},S^{q})\frac{\exp{\sum_{k=1}^{K}\theta
_{k}F_{k}(X^{q},S^{q})}}{Z(X)}\\
&=\sum\nolimits_{S^{q}}F_{k}(X^{q},S^{q})P(S^{q}|X^{q};\theta )\\
&=E_{P(S^{q}|X^{q};\theta )},
 \end{split}
\end{equation}
where $E_{P(S^{q}|X^{q};\theta )}[F_{k}(X^{q},S^{q})]$ is expected
feature value.

We can conclude that Equation (15) is strictly convex, so minimizing the
negative log-likelihood with gradient-descent algorithm can converge
to global minimum. In this paper, we use L-BFGS \cite{re39} for
parameters optimization. Furthermore, the computation of
$E_{P(S^{q}|X^{q};\theta )}[F_{k}(X^{q},S^{q})]$ is based on
marginal probability $p(y',y |X)$, which is defined as

\begin {equation}
\scalebox{0.85}{$
\begin{split}
 E_{P(S^{q}|X^{q};\theta
)}[F_{k}(X^{q},S^{q})]&=\sum
\nolimits_{S}F_{k}(X,S)P(S|X;\theta )\\
&=\sum_{i=1}^{n}\sum_{y',y\in Y}\theta
_{k}f_{k}(y',y,X)p(y',y|X).
\end{split}$}
\end{equation}

Similar to Semi-CRFs, the forward-backward method is employed to
compute $p(y',y|X)$. We use $\alpha(i,y)$ to denote the sum of
scores for all possible segments that end at position $i$  and
labels $y$. $\beta(i,y)$ denotes the sum of scores for all possible
segments from position $i+1$ whose previous segment ends in position
$i$ with label $y$.
\begin{equation}
\scalebox{0.9}{$\displaystyle{ \alpha
(i,y)=\sum_{d=1}^{L}\sum_{y'\in Y}\alpha
(i-d,y')\exp{(\sum_{k=1}^{K}\theta _{k}F_{k}(X,S))} \quad 0<i\leq n}
$}.
\end{equation}

\begin{equation}
\scalebox{0.9}{$ \displaystyle{\beta
(i,y)=\sum_{d=1}^{L}\sum_{y''\in Y}\beta
(i+d,y'')\exp{(\sum_{k=1}^{K}\theta _{k}F_{k}(X,S))} \quad 0<i<n
}$}.
\end{equation}
where $y'$ is the label of previous segment that ends in position
$i-d$ and $\alpha(0,y)=1$. $y''$ denotes the label of current
segment starts from position $i+1$ and $\beta(n,y) = 1$.

Based on the above discussion, the marginal probability
$p(y',y|X)$ can be computed as
\begin{equation}
p(y',y|X)=\frac{1}{Z(X)}\alpha(i-d,y')\exp{(\sum_{k=1}^{K}\theta_{k}F_{k}(X,S))\beta(i,y)}.
\end{equation}

\subsection{Constrained Viterbi decoding}

Decoding in DM-SMCRFs is to find an optimal segmentation and label assignment for a new sequence, which can be achieved by the globally most likely assignment.
\begin{equation}
S^{*}=arg\max \nolimits_{S}P(S|X;\theta).
\end{equation}

Traditional Viterbi algorithm defines a maximum unnormalized probability $V(i,y)$ of a segment that ends at position $i$ and labels $y$.
\begin{equation}
V(i,y)=\max\limits_{d=1,..,L}\max\limits_{y'}V(i-d,y')\phi
(y',y,X,i-d+1,i),
\end{equation}
where
\begin{equation}
\begin{aligned}
\phi (y',y,X,i-d+1,i)=&\exp \left \{ \sum_{k=1}^{K_{0}}\theta _{k}B_{k}(X^{q},S^{q})\right.\\
            &\left. + \sum_{{k}'=1}^{K_{1}}\theta _{{k}'}T_{{k}'}(X^{q},S^{q})\right.\\
            &\left. + \sum_{{k}''=1}^{K_{2}}\theta
            _{{k}''}G_{{k}''}(X^{q},S^{q})\right\}.
\end{aligned}
\end{equation}
The best segmentation and label assignment can be obtained by
tracing the path $\max\nolimits_{y',d=1,...,L}V(X,y)$.

Viterbi decoding algorithm for semi-Markov based models suffers from high computational cost as the complexity is proportional to $L$. In this paper, we proposed a constrained Viterbi decoding which combines two constraints from parsing and duration information. By pruning the subpaths that are impossible to lead to the best path, the decoding complexity of DM-SMCRFs can be reduced. 

First, we incorporate the constraint that most of manually assigned keyphrases are noun phrases in the decoding process, which forces the subsequence that is not a noun phrase to only pass through the state NKP. For example, when decoding the sentence "Automatic keyphrase extraction has received a lot of attention", as shown in Figure 2, "of attention" will not be assigned the label "KP" while "lot of attention" can be tagged with "KP" or "NKP".  The constraint defines the optimal subpath is 

\begin{equation}
V(i,d,y)=V(i,d,NKP),PT_{X(i-d+1,i))}\in [N-NP]
\end{equation}
where $PT_{X(i-d+1,i))}$ refers to the parse tree feature for the segment $X(i-d+1,i)$ and N-NP denotes the non-noun phrase.

Second, by taking the advantage of convexity of duration feature, subpaths that have no chance to result in the best predecessor of state KP can be pruned out. Let suppose $i_{th}$ element tagged with KP can achieve its maximum score via two possible paths $P_{1}$ and $P_{2}$ which are denoted in Figure 9 (a). For convenience, we use $y_{1}$ and $y_{2}$ to denote the state KP (keyphrase) and NKP (non-keyphrase) respectively. The paths can be expressed as

\begin{equation}
\left\{
\begin{array}{rcl}
V_{P1}(i,y_{1})=V(\tau _{1},y')\phi (y',y_{1},X,\tau_{1}+1,i),\\
V_{P2}(i,y_{1})=V(\tau _{2},y')\phi (y',y_{1},X,\tau_{2}+1,i),
\end{array} \right.
\end{equation}
where $y'\in\left \{ y_1,y_2 \right \}$.

If $P_{1}$  is more favorable than $P_{2}$, we can derive that
\begin{equation}
\frac{V_{P1}(i,y_{1})}{V_{P2}(i,y_{1})}>1.
\end{equation}

When extending to next $(i+1)_{th}$ element tagged with KP, the
ratio of path probability is
\begin{equation}
\begin{split}
\frac{V_{P1}(i+1,y_{1})}{V_{P2}(i+1,y_{1})} & = \frac{V(\tau
_{1},y')\phi (y',y_{1},X,\tau_{1}+1,i+1)}{V(\tau _{2},y')\phi
(y',y_{1},X,\tau_{2}+1,i+1)}\\
&=\frac{V_{P1}(i,y_{1})}{V_{P2}(i,y_{1})}*\frac{\phi
(y',y_{1},X,\tau_{2}+1,i)}{\phi (y',y_{1},X,\tau_{1}+1,i)}\\
&*\frac{\phi (y',y_{1},X,\tau_{1}+1,i+1)}{\phi
(y',y_{1},X,\tau_{2}+1,i+1)}.
\end{split}
\end{equation}

In our experiments, duration function tends to produce higher score
than that of observation function. The ratio of weight $\frac{\theta
_{i+1-\tau _{1}}^{y_{1}}}{\theta _{i+1-\tau
_{2}}^{y_{1}}}\frac{\theta _{i-\tau _{2}}^{y_{1}}}{\theta _{i-\tau
_{1}}^{y_{1}}}$ are always approximately equal to 1. We ignore these
insignificant contributions to the score $V(i,y)$. So
$\frac{V_{P1}(i+1,y_{1})}{V_{P2}(i+1,y_{1})}$ can be redefined as
\begin{equation}
\frac{V_{P1}(i+1,y_{1})}{V_{P2}(i+1,y_{1})}=\frac{V_{P1}(i,y_{1})}{V_{P2}(i,y_{1})}\frac{\exp{[D_{i+1-\tau
_{1}}^{y_{1}}]}}{\exp{[D_{i+1-\tau
_{2}}^{y_{1}}]}}\frac{\exp{[D_{i-\tau
_{2}}^{y_{1}}]}}{\exp{[D_{i-\tau _{1}}^{y_{1}}]}}.
\end{equation}

Let $d_{1}=i-\tau_{1}$, $d_{2}=i-\tau_{2}$, then $i+1-\tau_{1}=d+1$.
If $P_{1}$ still keeps its superiority, it should meet
\begin{equation}
\frac{\exp{[D_{d_{1}+1}^{y_{1}}]}}{\exp{[D_{d_{2}+1}^{y_{1}}]}}\frac{\exp{[D_{d_{2}}^{y_{1}}]}}{\exp{[D_{d_{1}}^{y_{1}}]}}\geq1.
\end{equation}

We define a new function $H(d)$ based on the natural logarithm of
$\exp{D_{d}^{y_{1}}}$, which is written as
\begin{equation}
H(d)=D_{d+1}^{y_{1}}-D_{d}^{y_{1}}.
\end{equation}

Deriving from duration feature defined in Section 4.2, we have $H'(d)\leq 0$. So it
can be concluded that
\begin{equation}
\frac{V_{P1}(i+1,y_{1})}{V_{P2}(i+1,y_{1})} >1, \quad if \quad
\tau_{1}\geq\tau_{2}.
\end{equation}

\begin{figure}
\centering
\scalebox{0.9}{
\includegraphics[width=4in,height = 2in ]{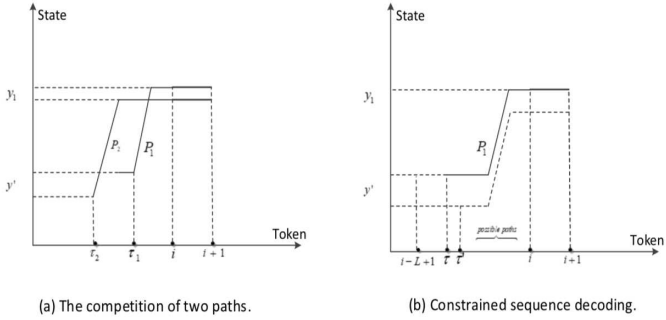}}
\caption {Sequence decoding. (a) The competition of two paths.(b) Constrained sequence decoding.} 
\end{figure}

Therefore if the best path for $V(i,y_{1})$ has arrived at $(\tau,{y}')$, then the best path for $V(i+1,y_{1})$ has arrived at $(\tau,{y}')$ and ${\tau}' > \tau$ from any state. As shown in Figure 9 (b), since the maximum length is $L$, $\tau$ should be in the range $(i-L+1,i)$ to allow the path is possible in $i$ and $i+1$. The corresponding constraint decoding is expressed as
\begin{equation}
\begin{aligned}
&V(i+1,y_{1})= \left \{ a,b \right \},\\
&a = V(\tau,y')\phi(y',y_{1},X,\tau+1,i+1),\\
&b = \max\limits_{\tau'}\max\limits_{y''}V(\tau',y'')\phi(y'',y_{1},X,\tau',i+1),\\
&s.t.  \; V(i,y_{1})= V(\tau,y')\phi(y',y_{1},X,\tau+1,i),\\
&\qquad i-L+1< \tau\leq i , \\
&\qquad \tau< \tau'\leq i+1.
\end{aligned}
\end{equation}

Combining the above two constraints, sequence decoding can reduce the average number of possible transitions between segments. The average lower bound of the segments assigned with state $y_{1}$ can be increased based on Equation (35), and the number of possible states for some segments is reduced by the hard constraint as shown in Equation (27). Further, the first constraint corrects the wrong assignment of labeling non-noun phrase as KP, which can improve the performance of keyphrase extraction.

\subsection{Time complexity}
Compared with CRFs, the inference and decoding time of semi-Markov based CRFs increases by a factor $L$ which is the maximum length of the segments. Table 3 lists the time complexity of CRFs, Semi-CRFs, Weak semi-CRFs and DM-SMCRFs, where $n$ is the sequence length and $\left | \mathcal{Y}\right |$ is the number of states. For the decoding complexity in DM-SMCRFs, we use $d^{*}$ to denote the average length of segment tagged with state KP, and $ \left | \mathcal{Y^{*}} \right |$ is the average number of possible states, then the complexity is $O\left [ n \left (  \left |  \mathcal{Y^{*}}\right |^{2}+L\left |  \mathcal{Y^{*}}\right |-  \left ( L-d^{*}\right ) \right ) \right ]$. It can be seen from Table 3 that DM-SMCRFs and Weak semi-CRFs are efficient than Semi-CRFs. Further, DM-SMCRFs reduce the decoding complexity compared with other semi-Markov based CRFs.

\begin{table}

\centering

\caption{The comparison of time complexity }
\scalebox{0.8}{
\begin{tabular}{lll} 

\hline
Model          & Inference & Decoding  \\ 
\hline
CRFs           &      $O(n\left |  \mathcal{Y}\right |^{2}) $         &        $O(n\left |  \mathcal{Y}\right |^{2}) $       \\
Semi-CRFs      &      $O(nL\left |  \mathcal{Y}\right |^{2})$        &     $O(nL\left |  \mathcal{Y}\right |^{2})$          \\
Weak semi-CRFs &     $O\left [ n \left (  \left |  \mathcal{Y}\right |^{2}+L\left |  \mathcal{Y}\right | \right ) \right ]$          &    $O\left [ n \left (  \left |  \mathcal{Y}\right |^{2}+L\left |  \mathcal{Y}\right | \right ) \right ]$             \\
DM-SMCRFs      &  $O\left [ n \left (  \left |  \mathcal{Y}\right |^{2}+L\left |  \mathcal{Y}\right | \right ) \right ]$          &     $O\left [ n \left (  \left |  \mathcal{Y^{*}}\right |^{2}+L\left |  \mathcal{Y^{*}}\right |-  \left ( L-d^{*}\right ) \right ) \right ]$            \\
\hline
\end{tabular}}
\end{table}

\section{Experimental Design}
In this section, we assess the performance of DM-SMCRFs via different duration modeling strategies, comparison with baselines and parameters sensitivity analysis (i.e. training size and maximal segment length). Furthermore, we demonstrate the decoding efficiency of constrained Viterbi algorithm compared with traditional Viterbi algorithm. Finally, we address the applicability of DM-SMCRFs by extracting keyphrases from long articles. Although we may not read these articles, extracted phrases can provide us an overview of the document.

\subsection{Datasets and Features}
We choose four datasets in the form of abstracts from scientific publications. Apart from Hulth dataset \cite{re10} that is related to Computer Engineering, we collect another three datasets concerned with History, Economics and Psycholgy. The datasets are summarized in Table 4.

Hulth dataset has two types of annotated keyphrases: uncontrolled keyphrases and controlled keyphrases. Following the experimental design by Hulth \cite{re10}, we choose uncontrolled keyphrases to measure the results of keyphrase extraction. For the other three datasets, we use author-specified keyphrases. In this paper, precision, recall and F1 score, the most common measures in information retrieval evaluation, are employed to evaluate the performance of keyphrase extraction. F1 score is chosen as final measure as it achieves a balance between precision and recall \cite{re10}\cite{re14}.

\renewcommand{\arraystretch}{1}
\begin{table*}
\centering \caption{Summary of datasets} \label{my-label}
\begin{tabular}{lllll}
\toprule[1pt]
Domain      & Source                            & \multicolumn{1}{c}{Time scale }               & \#Abstracts/\#keyphrases & Average length of keyphrases  \\
\midrule
Engineering & Inspec database(Hulth)            & \textbackslash{}          & \multicolumn{1}{c}{1500/14691}             & \multicolumn{1}{c}{2.33 }                        \\
History     & Journal of Eastern African Studies & 2010\textasciitilde{}2017 & \multicolumn{1}{c}{297/1672}              &\multicolumn{1}{c}{1.45}                       \\
Economics   & Journal of International Economics & 2013\textasciitilde{}2017 & \multicolumn{1}{c}{357/1531}                & \multicolumn{1}{c}{2.06}                        \\
Psychology  & Psychology Science                  & 2015\textasciitilde{}2017 & \multicolumn{1}{c}{337/1947 }               & \multicolumn{1}{c}{1.70 }                         \\
\bottomrule[1pt]
\end{tabular}
\end{table*}

For supervised keyphrase extraction, we use the following two types of features.

1) Syntactic features. Based on pre-specified noun groups pattern and assigned part-of-speech tags, we use a Boolean value to indicate whether the word is noun or not. Semi-Markov based models are operated on segment-level, the feature \{isNounPhrase\} is the sum of Boolean value of words within the segment.

2) Statistical features. First, we design the feature \{isInTile\}\cite{re14}. Each word is assigned a Boolean value to indicate whether it occurs in title or not. When extending to segment-level, the feature \{isInTitle\} is the sum of Boolean value of words within the segment. Specifically, for a segment confirming the pattern of noun groups, only a part of words occur in title, we define the corresponding feature \{isInTile\} is the size of this segment. Second, for semi-Markov based models, we use phrase length as another statistical feature.

\subsection{Baseline algorithms}
We use the following four algorithms to extract keyphrases as
baselines.

Naive Bayes \cite{re29}: The earliest supervised method applied in
keyphrase extraction is Naive Bayes, which takes advantage of Bayes'
theorem and still works even if the independence assumption does not
hold.

CRFs \cite{re14}: CRFs have been investigated for keyphrase
extraction in recent years and it is widely used in NLP tasks, for
example, POS tag and Named entity recognition.

Semi-CRFs \cite{re16}: Semi-CRFs are proposed to exploit
segment-level labels which are common in NP chunking and Named
entity recognition. For these tasks, Semi-CRFs always outperform
CRFs.

Weak semi-CRFs \cite{re15}: Weak semi-CRFs can improve inference
efficiency of Semi-CRFs by determining the length of next segment
and state separately. This model has been successfully applied in NP
chunking.

\subsection{Different duration modeling strategies}

To demonstrate the performance of different duration modeling
strategies, we use Gamma-like and Gaussian-like duration feature for
all datasets separately. Following the standard split
in Hulth dataset \cite{re10}, we set training data: testing data =
2: 1 for all datasets. The maximal segment length is set to 2, which
is a general size of keyphrases. As shown in Table 5, we have the
following observations:

1) For Engineering dataset, DM-SMCRFs with Gaussian-like duration
feature outperform that with Gamma-like duration feature in
precision, recall and F1 score.

2) For History dataset, DM-SMCRFs with Gamma-like duration feature
have advantages in recall and F1 score.

3) For Economics dataset, DM-SMCRFs with Gamma-like duration feature
have higher recall and F1 score while DM-SMCRFs with Gaussian-like
duration feature have higher precision.

4) For Psychology dataset, DM-SMCRFs with Gamma-like duration
feature achieve higher precision and F1 score while DM-SMCRFs with
Gaussian-like duration feature have higher recall.

\renewcommand{\arraystretch}{1.5}
\begin{table*}[h]
\centering \caption{Performance of DM-SMCRFs with different duration
feature} \label{my-label}
\begin{tabular}{|c|c|c|c|c|c|c|c|c|c|c|c|c|}
\hline \multirow{2}{*}{Duration feature} &
\multicolumn{3}{c|}{Engineering} & \multicolumn{3}{c|}{Psychology} &
\multicolumn{3}{c|}{Economics} & \multicolumn{3}{c|}{History} \\
\cline{2-13}
                                  & P         & R         & F1       & P         & R        & F1       & P        & R        & F1       & P        & R       & F1      \\ \hline
Gamma-like                        & 49.50     & 77.54    & 60.42 &
\textbf{45.48} & 40.21    & \textbf{42.69}    & \textbf{44.06}    &
\textbf{73.38} & \textbf{55.06} & 43.35 & \textbf{51.60}   & \textbf{48.11}   \\
\hline Gaussian-like & \textbf{51.11} & \textbf{79.53}     &
\textbf{62.23} & 28.27 & \textbf{42.94} & 34.09
& 38.66    & 61.65    & 47.52    & 42.09    & 31.42   & 35.98   \\
\hline 
\end{tabular}
\end{table*}

\renewcommand{\arraystretch}{1.5}
\begin{table*}[]
\centering
\caption{An example of keyphrase extraction in Engineering dataset}
\scalebox{0.8}{
\begin{tabular}{cl} 
\hline
Method                                                                & \multicolumn{1}{c}{The sentence with assigned keyphrases}                                                                                                                                                                                                                                                                                                                                                                                                                        \\ 
\hline
Manually Assigned Keyphrases                                          & \begin{tabular}[c]{@{}l@{}}The \textbf{RF field intensity~distribution} in the \textbf{human brain} becomes inhomogeneous due to wave behavior at \textbf{high field}. \\This is further complicated by the spatial distribution of \textbf{RF fieldpolarization} that must be considered to predict \\\textbf{imageintensity distribution}.\\An additional layer of complexity is involved when a \textbf{quadraturecoil }is used for transmission and reception.\end{tabular}  \\ 
\hline
\begin{tabular}[c]{@{}c@{}}DM-SMCRFs(Gamma-like)\\L=2\end{tabular}    & \begin{tabular}[c]{@{}l@{}}The \textbf{RF field} intensity~distribution in the \textbf{human brain} becomes inhomogeneous due~to wave behavior at high field.~\\This is further~complicated by the spatial distribution of \textbf{RF field~polarization} that must be considered to predict~\\\textbf{image~intensity} distribution.\\An additional layer of complexity is involved when a quadrature~\textbf{coil }is used for transmission and reception.\end{tabular}        \\ 
\hline
\begin{tabular}[c]{@{}c@{}}DM-SMCRFs(Gaussian-like)\\L=2\end{tabular} & \begin{tabular}[c]{@{}l@{}}The \textbf{RF field intensity~distribution} in \textbf{the human} brain becomes inhomogeneous due~to wave behavior at \textbf{high field}.~\\This is further~complicated by the spatial distribution of \textbf{RF field~polarization} that must be considered to predict~\\image~intensity distribution.\\An additional layer of complexity is involved when \textbf{a quadrature~coil }is used for transmission and reception.\end{tabular}        \\
\hline
\end{tabular}}
\end{table*}

We provide an example of keyphrase extraction in Engineering dataset. Given a paragraph, Table 6 shows manually assigned keyphrases and decoding results of DM-SMCRFs with Gamma-like and Gaussian-like duration features. We can find that DM-SMCRFs realize segment-level label assignment, such as keyphrase tag of \textit{RF field intensity distribution}. However, different duration modeling methods influence the performance of determining the size of keyphrase. For example, DM-SMCRFs (Gaussian-like)
dentify \textit{quadrature coil} as keyphrase while DM-SMCRFs (Gamma-like) only catch the word \textit{coil}.  From Table 5 and Table 6, we can conclude that different duration modeling strategies do affect the performance of keyphrase extraction. Generally, DM-SMCRFs with Gamma-like duration feature tend to outperform that with Gaussian-like duration feature. For the following sections, we use Gaussian-like duration feature in Engineering dataset and Gamma-like duration feature in the other three datasets.

\subsection{Comparing with baselines}

To evaluate the effectiveness of proposed model, we choose the baselines presented in Section 5.2 for comparative study. For semi-Markov based models, the maximal segmental length is set to 2. Table 7 shows the performance of five methods for keyphrase extraction, we observe that

1) For all selected datasets, DM-SMCRFs can achieve the best F1 score, which make an approximate increase of $10\%$ compared with baselines. Besides, in Engineering, Economics and History dataset, DM-SMCRFs produce higher recall and F1 score than baselines. 

2) For the comparison between semi-Markov based models, Semi-CRFs can achieve the highest precision in Psychology, Economics and History dataset. However, Weak semi-CRFs have no advantage in pre-specified measures for keyphrase extraction. 

Generally, semi-Markov based models tend to produce redundant segments labeled with keyphrases, which may results in the low precision compared with Naive Bayes and CRFs. Moreover, these selected four datasets comes from different disciplines. Based on the same linguistic and statistical features, the performance of a classifier (e.g. CRFs and Semi-CRFs) can change significantly by varying dataset.

\begin{table*}[]
\centering \caption{Performance of baselines and DM-SMCRFs}
\label{my-label}
\begin{tabular}{|c|c|c|c|c|c|c|c|c|c|c|c|c|}
\hline \multirow{2}{*}{Methods} & \multicolumn{3}{c|}{Engineering} &
\multicolumn{3}{c|}{Psychology} & \multicolumn{3}{c|}{Economics} &
\multicolumn{3}{c|}{History} \\ \cline{2-13}
                         & P         & R         & F1       & P         & R        & F1       & P        & R        & F1       & P        & R       & F1      \\ \hline
Naive Bayes              & 57.95     & 41.83     & 48.59    & 49.50
& 32.49    & 39.23    & 82.42    & 30.49    & 44.51 & \textbf{57.11}
& 37.50 & 45.28   \\ \hline CRFs                     & \textbf{85.61
}    & 33.03 & 47.67    & 27.22     & \textbf{46.06 }   & 34.22    &
\textbf{82.60}    & 30.20    & 44.23    & 51.36    & 43.71   & 47.22   \\
\hline Semi-CRFs                & 27.41     & 72.61     & 39.79    &
\textbf{76.48} & 22.15    & 34.35    & 73.54    & 40.01    & 51.82 &
50.88 & 17.13   & 25.63   \\ \hline Weak semi-CRFs           & 37.77
& 65.99     & 48.04    & 32.34     & 18.60    & 23.62    &
41.21    & 57.38    & 47.97    & 42.60    & 46.69   & 44.55   \\
\hline DM-SMCRFs                & 51.11     & \textbf{79.53 }    &
\textbf{62.23} & 45.48 & 40.21    & \textbf{42.69}    & 44.06    &
\textbf{73.38}    & \textbf{55.06}    & 43.35    & \textbf{51.60}   & \textbf{48.11}   \\
\hline
\end{tabular}
\end{table*}

\subsection{Influence of the maximal segment length}
In this section, we assess how the different choices of the maximal
segment length $L$ on semi-Markov based models affect the
performance of keyphrase extraction.

We choose $L=2$ and $L=3$ based on the observation of manually
assigned  keyphrases. Figure 10 shows the performance of DM-SMCRFs
with different segment length. We can see that DM-SMCRFs with $L=2$
can achieve higher precision and F1 score. When $L$ is set to 3, we
compare the performance of three semi-Markov based models, as shown
in Figure 11. For Engineering and Psychology datasets, DM-SMCRFs show
its superiority in F1 score. Semi-CRFs with $L=3$ can outperform the
other two models in precision for four datasets.

Table 8 lists all the results of three semi-Markov based models.
From this table we can find that increasing the maximal segment
length in semi-Markov based models degrades the performance of
keyphrase extraction. In most cases, manually assigned keyphrases
are the combination of two words. When $L$ is greater than 2,
semi-Markov based models are more likely to produce segments with
larger size, resulting in redundancy and wrong segmentation of
keyphrases.

\begin{figure}[]
\centering
\includegraphics[width=3.5in,height = 2.9in]{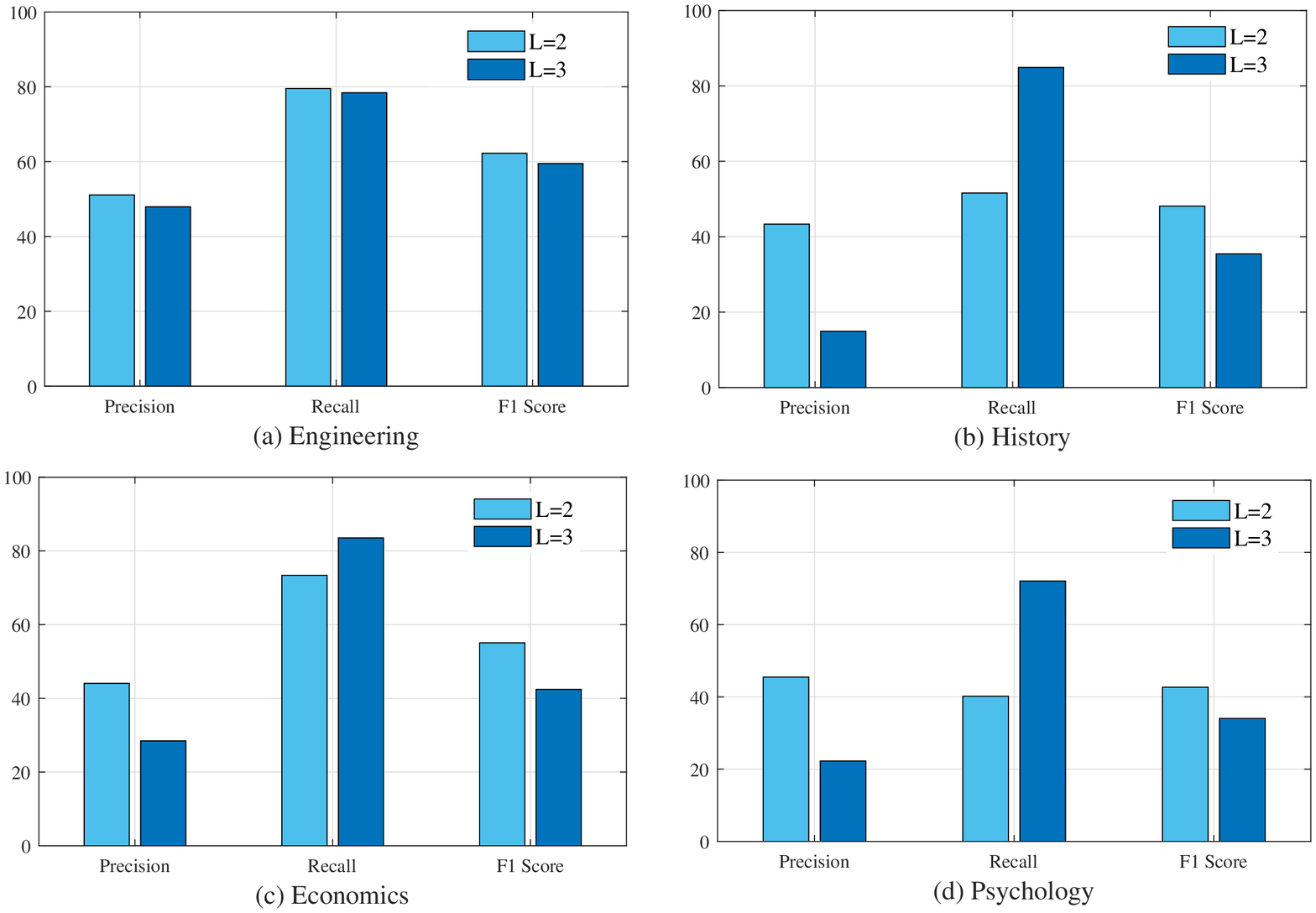}
\caption {Performance of DM-SMCRFs with different maximal segment
length.(a) Engineering.(b) History.(c) Economics.(d) Psychology.}
\label{fig:secondfigure}
\end{figure}

\begin{figure}[]
\centering
\includegraphics[width=3.5in,height = 2.9in]{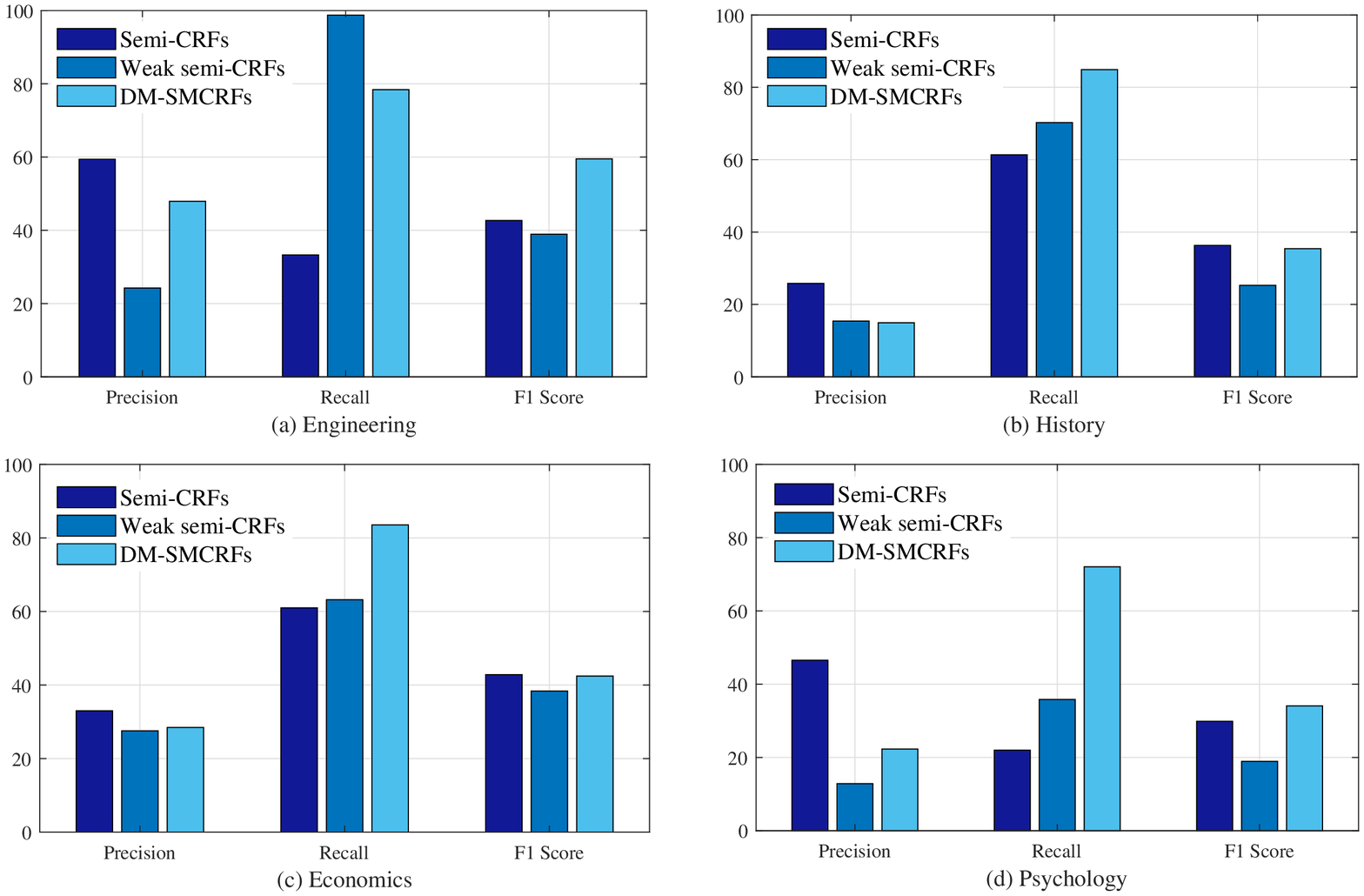}
\caption {Performance of semi-Markov based models with $L=3$.(a)
Engineering.(b) History.(c) Economics.(d) Psychology.}.
\label{fig:secondfigure}
\end{figure}

\renewcommand{\arraystretch}{1}
\begin{table*}[]
\centering \caption{Performance of semi-Markov based models with
different maximal segment length} \label{my-label}
\begin{tabular}{|c|c|c|c|c|c|c|c|c|c|c|}
\hline \multirow{2}{*}{Dataset}     & \multirow{2}{*}{Segment
length} & \multicolumn{3}{c|}{Semi-CRFs} & \multicolumn{3}{c|}{Weak
semi-CRFs} & \multicolumn{3}{c|}{DM-SMCRFs} \\ \cline{3-11}
                             &                                 & P        & R        & F1       & P          & R          & F1        & P        & R        & F1       \\ \hline
\multirow{2}{*}{Engineering} & L=2                             &
27.41    & 72.61    & 39.79    & 37.77      & 65.99      & 48.04 &
51.11    & \textbf{79.53}    & \textbf{62.23 }   \\ \cline{2-11}
                             & L=3                             & 59.42    & 33.29    & 42.67    & 24.25      & 98.77      & 38.94     & 44.18    & 41.43   & 51.40    \\ \hline
\multirow{2}{*}{History}     & L=2                             &
50.88    & 17.13    & 25.63    & 42.60      & 46.69      & 44.55 &
43.35    & 51.60    &\textbf{ 48.11 }   \\ \cline{2-11}
                             & L=3                             & 25.81    & 61.36    & 36.33    & 15.42      & 70.27      & 25.29     & 14.96    & \textbf{84.90}    & 35.44    \\ \hline
\multirow{2}{*}{Economics}   & L=2                             &
73.54    & 40.01    & 51.82    & 41.21      & 57.38      & 47.97 &
44.06    & 73.38    &\textbf{ 55.06 }   \\ \cline{2-11}
                             & L=3                             & 33.00    & 60.98    & 42.83    & 27.53      & 63.22      & 38.35     & 28.45    & \textbf{83.53  }  & 42.44    \\ \hline
\multirow{2}{*}{Psychology}  & L=2                             &
76.48    & 22.15    & 34.35    & 32.34      & 18.60      & 23.62 &
45.48    & 40.21    & \textbf{42.69}    \\ \cline{2-11}
                             & L=3                         & 46.52    & 21.97    & 29.85    & 12.85      & 35.81      & 18.91     & 22.29    & \textbf{72.05}    & 34.05    \\ \hline
\end{tabular}
\end{table*}

\subsection{Influence of training size}
To demonstrate the effect of training size, we set the training size from 50\% and 90\% of total dataset. The choices of duration feature and segment length are the same as the setting in Section 5.4. Figure 12, 13, 14 and 15 show the performance with different training size on four datasets. Table 9 and 10  list F1 score with different training size. The observations are made as follows:

1) For baselines and DM-SMCRFs, they are insensitive to the change of training size as the values of three measures (i.e.precision, recall and F1 score) vary within a small range. For example, in Engineering dataset, with the increase of training size, the precision fluctuates within $61.27\pm 1.80$.

2) For Engineering and Economics datasets, DM-SMCRFs can outperform baselines in recall and F1 score. Generally, DM-SMCRFs achieve higher F1 score than baselines.

\renewcommand{\arraystretch}{1}
\begin{table*}[]
\centering \caption{F1 score of baselines and DM-SMCRFs in
Engineering and History datasets} \label{my-label}
\begin{tabular}{|c|c|c|c|c|c|c|c|c|c|c|}
\hline \multirow{2}{*}{Methods} & \multicolumn{5}{c|}{Engineering} &
\multicolumn{5}{c|}{History}
\\ \cline{2-11}
                         & 50\%           & 60\%           & 70\%           & 80\%           & 90\%           & 50\%          & 60\%           & 70\%           & 80\%           & 90\%           \\ \hline
Naive Bayes              & 48.94          & 49.11          & 48.49 &
48.38          & 51.56          & 46.35         & 46.45          &
45.22          & 45.46          & 52.40          \\ \hline CRFs &
47.08          & 54.77          & 47.00          & 47.64 & 47.38
& \textbf{52.0} & 46.83          & \textbf{46.63} & \textbf{46.32} &
53.63          \\ \hline Semi-CRFs                & 39.73          &
39.78          & 40.00          & 36.45          & 40.43          &
23.49         & 24.43          & 24.75          & 22.42          &
23.90          \\ \hline Weak semi-CRFs           & 52.35          &
50.87          & 50.61          & 48.80          & 48.19          &
44.83         & 43.92          & 43.44          & 44.10          &
50.77          \\ \hline DM-SMCRFs                & \textbf{61.72} &
\textbf{58.13} & \textbf{62.30} & \textbf{61.63} & \textbf{62.56} &
47.95         & \textbf{46.95} & 45.95          & 44.90          &
\textbf{53.80} \\ \hline
\end{tabular}
\end{table*}

\renewcommand{\arraystretch}{1}
\begin{table*}[h]
\centering \caption{F1 score of baselines and DM-SMCRFs in Economics
and Psychology datasets} \label{my-label}
\begin{tabular}{|c|c|c|c|c|c|c|c|c|c|c|}
\hline \multirow{2}{*}{Methods} & \multicolumn{5}{c|}{Economics} &
\multicolumn{5}{c|}{Psychology}
\\ \cline{2-11}
                         & 50\%           & 60\%           & 70\%           & 80\%           & 90\%           & 50\%           & 60\%           & 70\%           & 80\%           & 90\%           \\ \hline
Naive Bayes              & 44.20          & 44.18          & 42.45 &
42.11          & 41.87          & 40.91          & 41.28 & 38.71
& 38.88          & 43.89          \\ \hline CRFs & 43.87          &
43.93          & 42.13          & 41.60 & 41.05          & 33.78
& 30.91          & 33.42 & 32.72          & 33.0           \\ \hline
Semi-CRFs & 50.54          & 51.11          & 51.55          & 52.64
& 51.89          & 33.62          & 34.83          & 34.10 & 34.50
& 35.67          \\ \hline Weak semi-CRFs & 43.46          & 44.90
& 47.32          & 46.39 & 47.69          & 24.51          & 24.89
& 23.28 & 24.27          & 26.06          \\ \hline DM-SMCRFs &
\textbf{53.37} & \textbf{53.73} & \textbf{55.40} & \textbf{57.45} &
\textbf{57.55} & \textbf{44.09} & \textbf{45.15} & \textbf{42.26} &
\textbf{42.90} & \textbf{43.87} \\ \hline
\end{tabular}
\end{table*}

\begin{figure}[]
\centering
\includegraphics[width=3.8in,height = 3.2in]{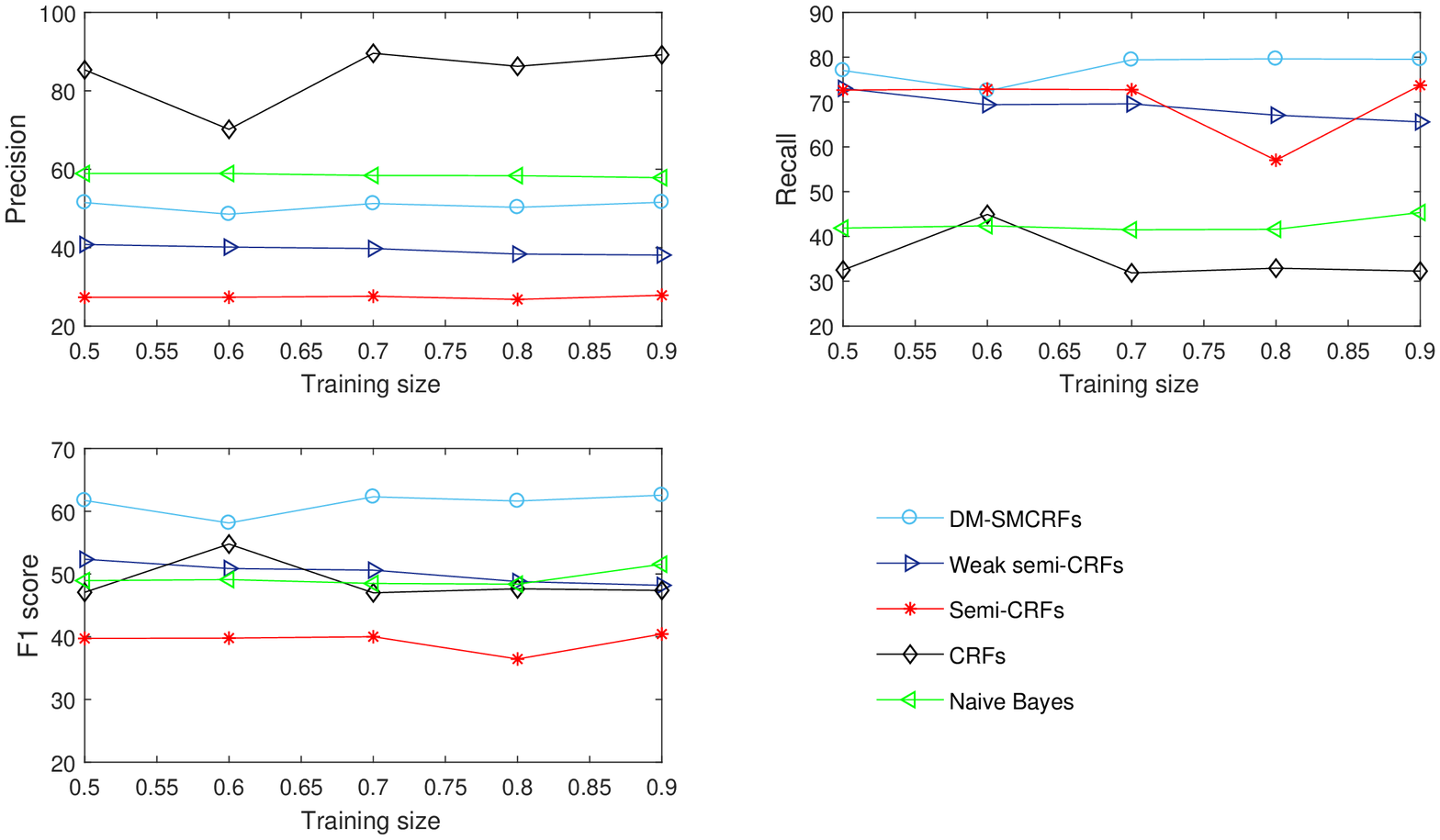}
\caption {Engineering: Performance with different training size. }
\label{fig:secondfigure}
\end{figure}

\begin{figure}[]
\centering
\includegraphics[width=3.8in,height = 3.2in]{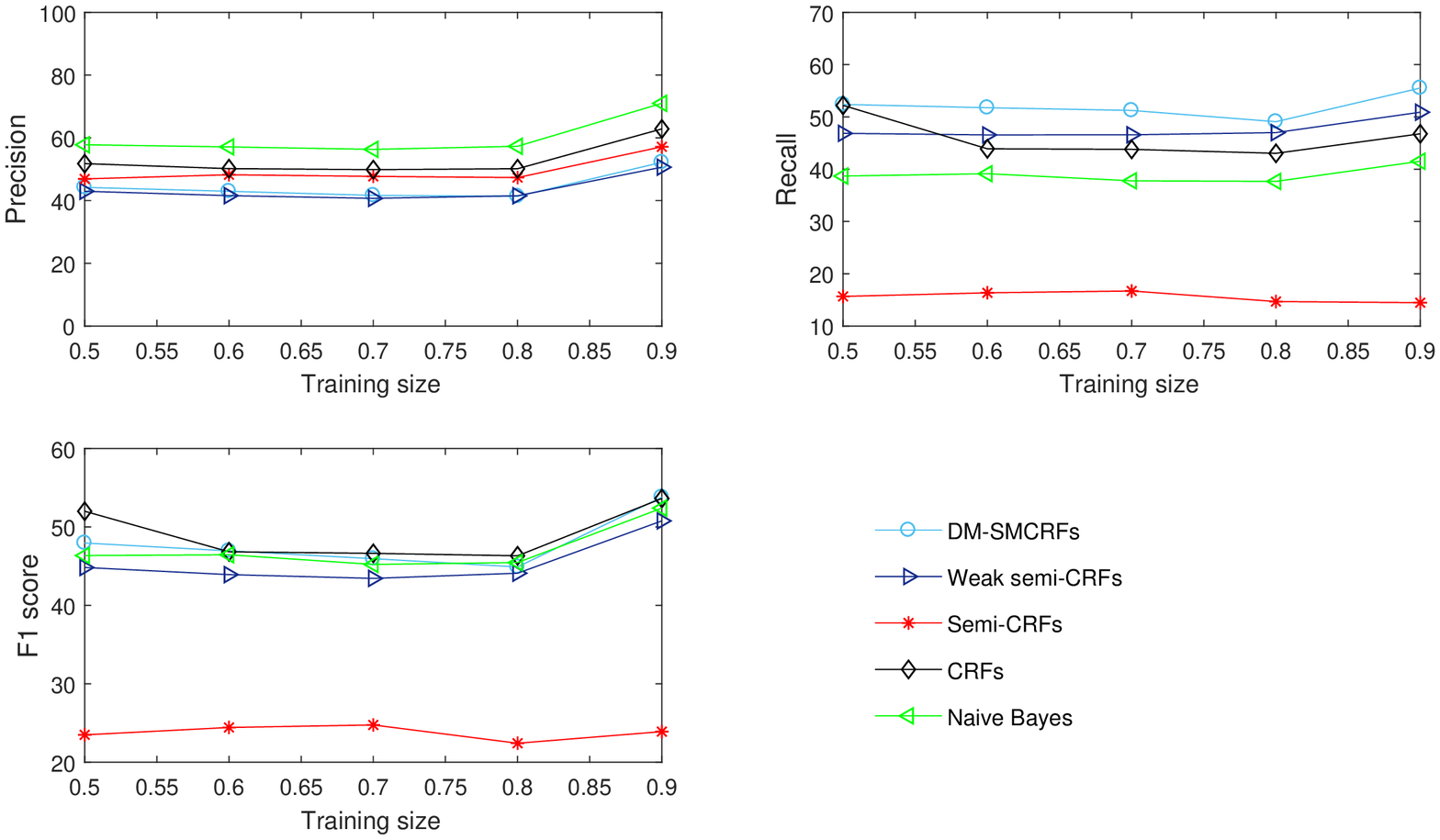}
\caption {History: Performance with different training size.  }
\label{fig:secondfigure}
\end{figure}

\begin{figure}[]
\centering
\includegraphics[width=3.8in,height = 3.2in]{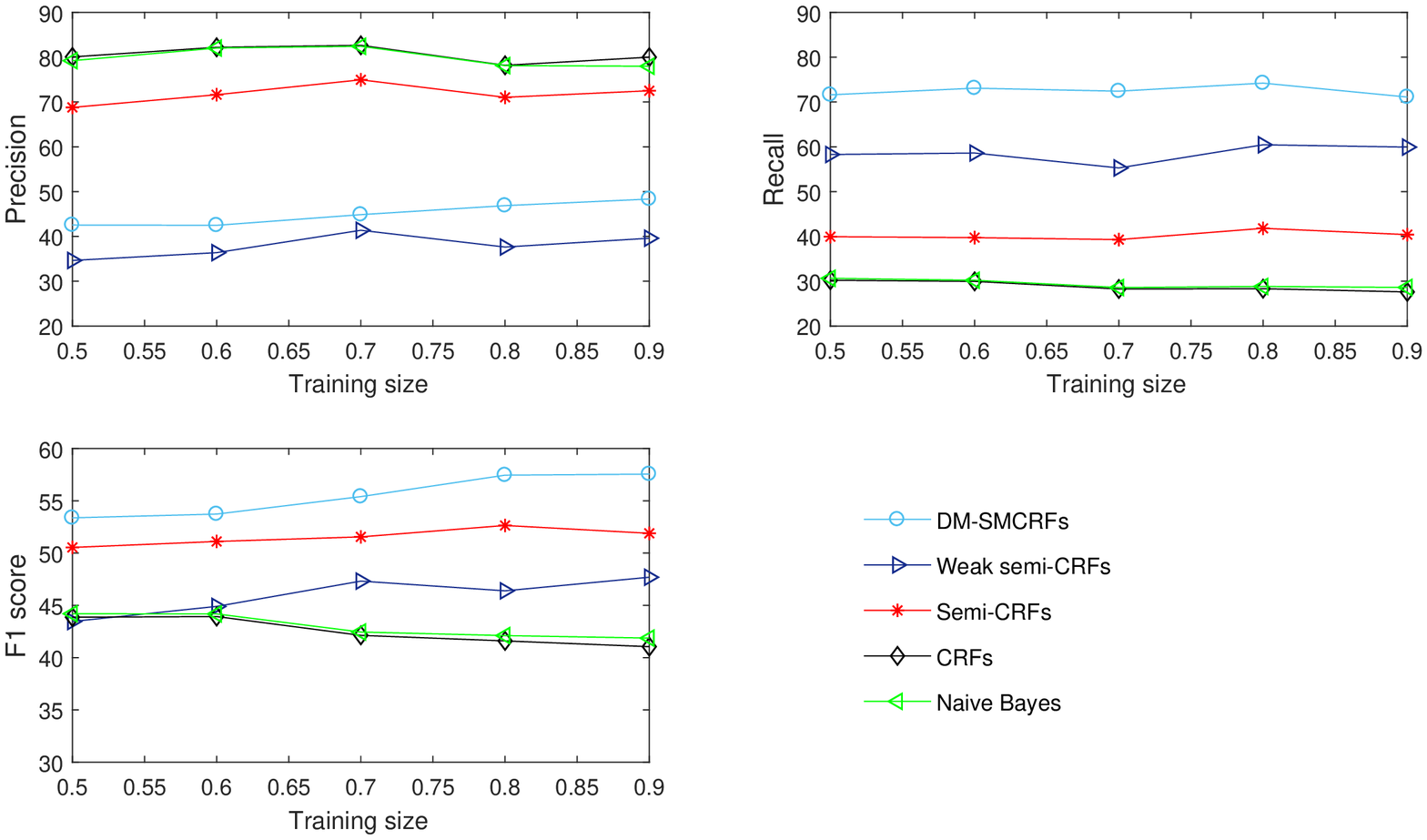}
\caption {Economics: Performance with different training size. }
\label{fig:secondfigure}
\end{figure}

\begin{figure}[]
\centering
\includegraphics[width=3.8in,height = 3.2in]{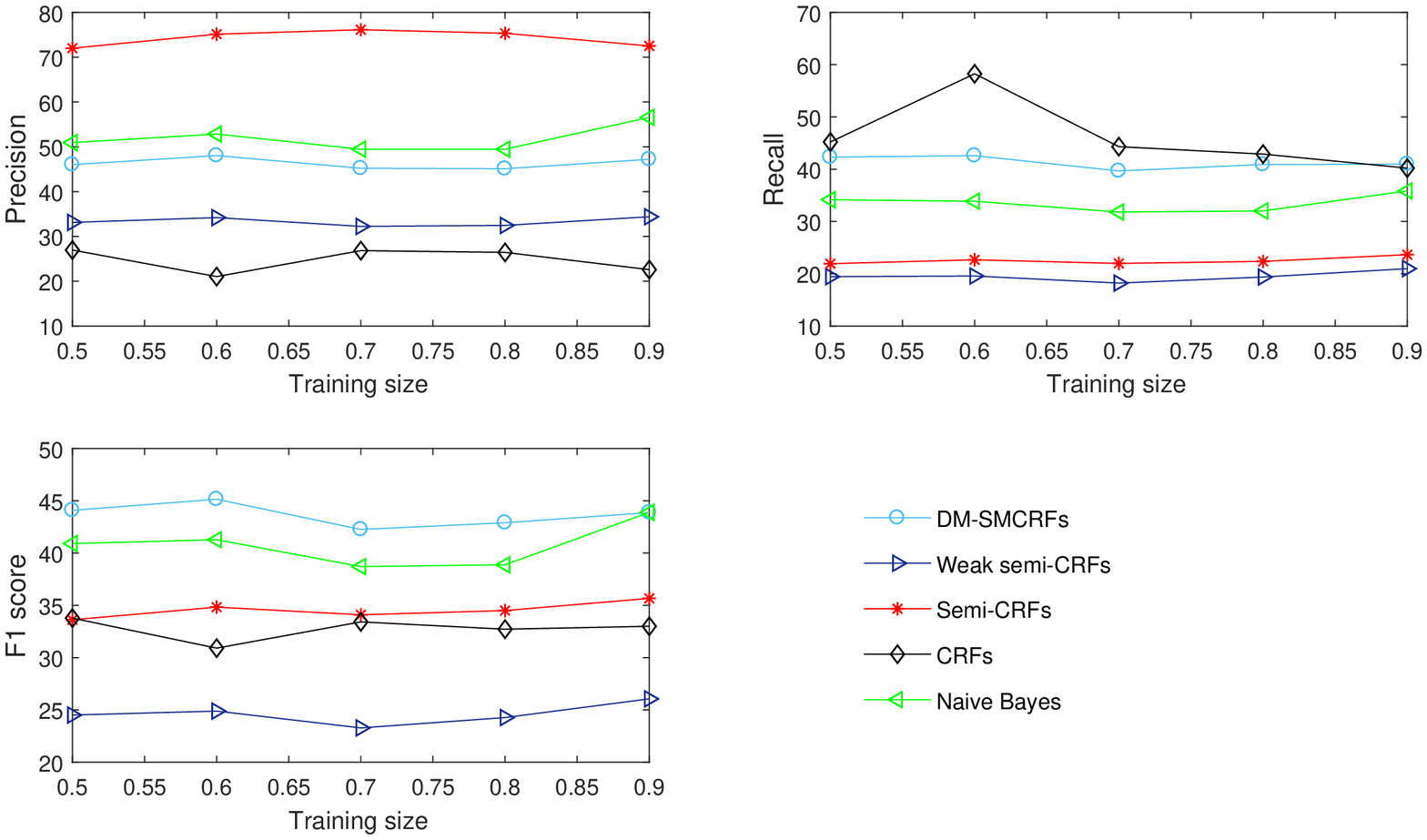}
\caption {Psychology: Performance with different training size. }
\label{fig:secondfigure}
\end{figure}

\subsection{Decoding efficiency}
In this section, we assess the effectiveness of proposed constrained Viterbi algorithm.

Due to limited size of collected data, we choose Engineering dataset to measure decoding time. Training size : test size = 2:1 and the maximal segment length is 2. Figure 16 shows the decoding time of constrained Viterbi algorithm and traditional Viterbi algorithm. It is obvious that the decoding efficiency is proportional to testing size and constrained Viterbi requires less time in sequence decoding. For keyphrase extraction, there are two states KP and NKP.  $\left | \mathcal{Y^{*}} \right |$ obtained from the experiments is usually within the range [1,1.8]. Similarly, $d^{*}$ in the experiments is close to the value of maximum length $L$ which is set to 2 or 3. Therefore the running time of the constrained Viterbi is slightly more efficient than the traditional Viterbi.

\begin{figure}[]
\centering
\includegraphics[width=3.5in,height = 2in]{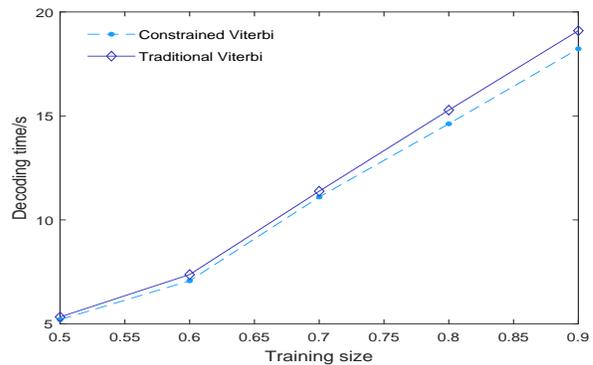}
\caption {Decoding time of constrained Viterbi and traditional
Viterbi. } \label{fig:secondfigure}
\end{figure}

We further demonstrate the decoding results of constrained Viterbi on four datasets compared with traditional Viterbi. Figure 17 visualizes the results on Engineering dataset and Table 11 lists all the results of datasets. In Figure 17, we observe that with different duration modeling strategies, constrained Viterbi algorithm can always achieve better performance than traditional Viterbi. From Table 11, we also find that constrained Viterbi algorithm can keep its superiority on four datasets. As presented in Section 4.4, the constrained Viterbi algorithm incorporates the designed hard constraint to correct the unlikely labeling of state KP to non-noun phrase, which improves the performance of keyphrase extraction.

\begin{figure}[]
\centering
\includegraphics[width=3.5in,height = 2in]{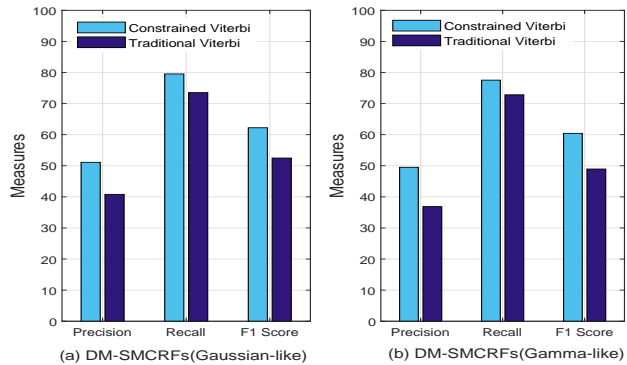}
\caption {Decoding results of constrainted Viterbi and traditional
Viterbi.(a) DM-SMCRFs(Gaussian-like). (b) DM-SMCRFs(Gamma-like).}
\label{fig:secondfigure}
\end{figure}

\begin{table}[h]
\centering \caption{Decoding results of four datasets}
\label{my-label2}
\begin{tabular}{cc|c|c|c}
\toprule[1pt]
                           & Decoding method & P & R & F1 \\
                           \midrule
\multirow{2}*{Engineering} & traditional Viterbi &40.80 & 73.51 & 52.48 \\
                           & constrained Viterbi & \textbf{51.12} & \textbf{79.53} &
                         \textbf{62.23} \\
\multirow{2}*{History} & traditional Viterbi & 35.83 & 47.43 & 40.82 \\
                       & constrained Viterbi & \textbf{43.35} & \textbf{51.60} & \textbf{48.11}
                       \\
\multirow{2}*{Economics} & traditional Viterbi &41.11 & 55.61 & 47.27 \\
                         & constrained Viterbi & \textbf{44.06} & \textbf{73.38} &
                         \textbf{55.06} \\
\multirow{2}*{Psychology} &traditional Viterbi & \textbf{46.10} & 36.36 & \textbf{40.66}\\
                          & constrained Viterbi & 45.48 & \textbf{40.21} &
                          \textbf{42.69} \\
\bottomrule[1pt]
\end{tabular}
\end{table}

\subsection{Application to long articles}

In practice, there are too many articles only containing titles and context. Extracting keyphrases from these articles provides a quick review of the document for readers. In this section, we choose two long articles\footnote{\scriptsize{https://thediplomat.com/2018/05/the-u-s-china-trade-war/, https://www.bbc.com/news/technology-44481510}} and one chapter of Bibles\footnote{\scriptsize{https://www.biblegateway.com/passage/?search=Matthew+14\&version=NIV}} to demonstrate the applicability of DM-SMCRFs. And we use the model trained in Engineering dataset to extract keyphrases.

As shown in Table 12, we list all the  keyphrases extracted from Matthew 14. Furthermore, keyphrases extracted from top 7 paragraphs in the second article and top 14 paragraphs for the third are presented. We find that DM-SMCRFs can extract informative phrases from the document and these keyphrases briefly describe the context of articles. Furthermore, extracted keyphrases can be treated as an extension of phrases in title. For example, AI in title and artificial intelligence in the context.

\renewcommand{\arraystretch}{1.5}
\begin{table}[h]
\centering \caption{Extracting keyphrases from long articles}
\scalebox{0.8}{
\begin{tabular}{ll}
\toprule[1pt]
Articles                                                                                                                       & Extracted keyphrases                                                                                                                                                                                                                                                                                                                                                                                                                                                                                                                                                                                                                                                               \\
\midrule[0.5pt]
\begin{tabular}[c]{@{}l@{}}Bible Matthew 14\\(New International Version)\\(710 words)\\\end{tabular}                          & \begin{tabular}[c]{@{}l@{}}John the Baptist, tetrarch, Jesus,\\miraculous powers, Herod, prison, \\Herodias, Philip¡s wife, John a prophet, \\oath, Baptist, platter, Five Thousand, \\boat,solitary place, crowd, compassion, \\remote place, five loaves, two fish, disciples,\\twelve basketfuls, broken pieces, water,\\lake,little faith, Son of God, Gennesaret.\end{tabular}                                                                                                                                                                                                                                                                                                \\
\midrule
\begin{tabular}[c]{@{}l@{}}The US-China Trade War~\\A simple plan: action,~\\confusion, and retreat\\(2369 words)\end{tabular} & \begin{tabular}[c]{@{}l@{}}Pyrrhic victory,Napoleon,\\General Mikhail Kutuzov,\\historian John Lewis Gaddis,\\subordinate to policy,Donald Trump,\\trade war,China,coherent strategy,\\battle,another round,enforcement actions,\\tariffs,Trump's apparent lead negotiator,\\U.S. retreat,U.S. trade policy,second round,\\surprising moratorium,high-level discussions,\\Chinese President Xi Jinping,ZTE Corporation,\\telecommunications,Ross,combined civil,\\Korea,export laws,probationary period,\\agreement,White House,denial order,fatal,\\30 percent,Intel,computers,main business,\\trade talks,national boundaries,\\trade negotiationsmaterial change.\end{tabular}  \\
\midrule[0.5pt]
\begin{tabular}[c]{@{}l@{}}AI gives silenced radio~\\journalist his voice back\\(868 words)\end{tabular}                       & \begin{tabular}[c]{@{}l@{}}radio journalist,voice,artificial intelligence,\\Media Group,rare neurological condition,\\technology company,Dupree,voice recordings,\\financial unknown,Mr Dupree,market,\\speech patterns,CereProc,neural networks,\\intelligence system,individual,tiny pieces,\\ordered sequence,each word,human brain.\end{tabular}                                                                                                                                                                                                                                                                                                                               \\
\bottomrule[1pt]
\end{tabular}}
\end{table}

\section{Conclusion}
In this paper, we propose a novel approach called duration modeling with semi-Markov Conditional Random Fields (DM-SMCRFs) to extract keyphrases from the document. Different from traditional methods for keyphrase extraction, DM-SMCRFs sequentially classify the phrase as keyphrase or non-keyphrase, which avoids preprocessing for candidate phrase generation. Further, DM-SMCRFs assume the independence between state transition and state duration to allow explicitly modeling the distribution of duration of keyphrase to further explore state duration information. This mechanism improves the performance of exact segmentation of keyphrases compared with other semi-Markov based models. Based on the convexity of parametric duration feature derived from duration distribution, a constrained Viterbi algorithm is derived to reduce the decoding complexity and improve the performance of DM-SMCRFs. Our obtained results show that constraint Viterbi algorithm can outperform the traditional Viterbi algorithm in efficiency and performance. Moreover, compared with baseline methods, DM-SMCRFs can achieve better performance in keyphrase extraction.

It should be noticed that feature \{isInTitle\} plays an important role in keyphrase extraction. However, it is difficult for DM-SMCRFs to encode this feature in some articles with abstract titles, such as \textit{Beyond Faces and Expertise}. For future work, we will consider to enrich global features in supervised keyphrase extraction. Meanwhile, in this paper we only consider three common duration feature functions for duration modeling. Designing a general duration feature function that can be applied to various domains may be useful for keyphrase extraction.

\section{Acknowledgements}
This work is supported by a project donated by Mr. MW Lau of CityU
project No. 9220083.

\bibliographystyle{ieeetr}
\bibliography{ref4}

\end{document}